\documentclass[prb,amsmath,twocolumn,amssymb]{revtex4-1}
\usepackage{times}
\usepackage{graphicx}
\usepackage{amsmath, braket}
\usepackage{amssymb}
\usepackage{natbib}
\usepackage{bm, color, ulem}

\def\lsim{\lower -0.3ex \hbox{$<$} \kern -0.75em \lower 0.7ex \hbox{$\sim$}}
\def\gsim{\lower -0.3ex \hbox{$>$} \kern -0.75em \lower 0.7ex \hbox{$\sim$}}

\begin{document}

\title{Direct measurement of discrete valley and orbital quantum numbers \\ in a multicomponent quantum Hall system}

\author{B.M. Hunt$^{1,2,3}$,  J.I.A. Li$^2$, A.A. Zibrov$^4$, L. Wang$^5$, T. Taniguchi$^6$, K. Watanabe$^6$, J. Hone$^5$, C. R. Dean$^2$, M. Zaletel$^7$, R.C. Ashoori$^1$, A.F. Young$^{1,4,*}$\\
\normalsize{$^{1}$Department of Physics, Massachusetts Institute of Technology, Cambridge, MA 02139}\\
\normalsize{$^{2}$Department of Physics, Columbia University, New York, NY 10027, USA}\\
\normalsize{$^{3}$Department of Physics, Carnegie Mellon University, Pittsburgh, Pennsylvania 15213, USA}\\
\normalsize{$^{5}$Department of Mechanical Engineering, Columbia University, New York, NY 10027, USA}\\
\normalsize{$^{6}$Advanced Materials Laboratory, National Institute for Materials Science, Tsukuba, Ibaraki 305-0044, Japan}\\
\normalsize{$^{7}$Station Q, Microsoft Research, Santa Barbara, California 93106-6105, USA}\\
\normalsize{$^{4}$Department of Physics, University of California, Santa Barbara CA 93106 USA}\\
\normalsize{$^\ast$andrea@physics.ucsb.edu}
}

\maketitle
\textbf{Strongly interacting two dimensional electron systems (2DESs) host a complex landscape of broken symmetry states.
The possible ground states are further expanded by internal degrees of freedom such as spin or valley-isospin.  While direct probes of spin in 2DESs were demonstrated two decades ago\cite{barrett_optically_1995}, the valley quantum number has only been probed indirectly in semiconductor quantum wells\cite{shkolnikov_observation_2005}, graphene mono-\cite{young_spin_2012,feldman_unconventional_2012} and bilayers\cite{feldman_broken-symmetry_2009, weitz_broken-symmetry_2010, zhao_symmetry_2010, maher_evidence_2013, velasco_jr_competing_2014, kou_electron-hole_2014, maher_tunable_2014, lee_chemical_2014, shi_energy_2016}, and, transition metal dichalcogenides.\cite{xu_spin_2014}
Here, we present the first direct experimental measurement of valley polarization in a two dimensional electron system\cite{young_capacitance_2011}, effected via the direct mapping of the valley quantum number onto the layer polarization in bilayer graphene at high magnetic fields.  We find that the layer polarization evolves in discrete steps across 32 electric field-tuned phase transitions between states of different valley, spin, and orbital polarization.  Our data can be fit by a model that captures both single particle and interaction induced orbital, valley, and spin anisotropies, providing the most complete model of this complex system to date. Among the newly discovered phases are theoretically unanticipated orbitally polarized states stabilized by skew interlayer hopping. The resulting roadmap to symmetry breaking in bilayer graphene paves the way for deterministic engineering of fractional quantum Hall states\cite{papic_topological_2014,apalkov_controllable_2010}, while our layer-resolved technique is readily extendable to other two dimensional materials where layer polarization maps to the valley or spin quantum numbers, providing an essential direct probe that is a prerequisite for manipulating these new quantum degrees of freedom.}

The single-particle energy spectrum of a 2DES in a large magnetic field collapses into Landau levels (LLs) containing $N^\Phi$ degenerate states, with  $N^\Phi$ the number of magnetic flux quanta penetrating the sample.  The width in energy of the LL bands is limited only by disorder, making electronic interactions effectively strong in a clean system even when their
absolute scale is weak.  The simplicity of the starting LL wavefunctions, combined with the high degree of control available in two dimensional electron systems, make LLs a promising venue for engineering electronic ground states based on electron correlations.  However, the difficulty of simulating interacting electron problems necessitates experimental input to constrain the possible ground states, particularly in the presence of internal degeneracy.
The Bernal bilayer graphene (B-BLG) zero energy Landau level (ZLL) provides an extreme example of such degeneracy.  In B-BLG, LLs $|\xi  N \sigma\rangle$ are labeled by their electron spin $\sigma = \uparrow,\downarrow$, valley $\xi = +,-$, and ``orbital'' index $N \in \mathbb{Z}$. Electrons in valley $+/-$ are localized near points $K / K'$ of the hexagonal Brillouin zone, while the index $N$ is closely analogous to the LL-index of conventional LL systems.
The energies of the LLs are approximately $\epsilon_{\sigma  \xi N} \approx \hbar \omega_c \textrm{sign}(N) \sqrt{ N (N-1)}$, where $\hbar \omega_c$ is the cyclotron energy, leading to an eight-fold nearly degenerate ZLL comprising the $N=0$ and $N=1$ orbitals and all possible spin and valley isospin combinations.

Resolving the order in which the eight components fill as electrons are added is one of the key open questions in the physics of bilayer graphene, and is essential to efforts to use bilayer graphene to engineer exotic phases of matter based on electronic correlations\cite{apalkov_stable_2011,papic_topological_2014}. Due to an approximate SU(4) symmetry relating spin and valley, determining which of the components are filled is non-trivial.
Past experiments\cite{feldman_broken-symmetry_2009, weitz_broken-symmetry_2010, zhao_symmetry_2010, bao_evidence_2012, maher_evidence_2013, velasco_jr_competing_2014, kou_electron-hole_2014, maher_tunable_2014, lee_chemical_2014, shi_energy_2016} have observed numerous phase transitions between gapped ground states at both integer \cite{weitz_broken-symmetry_2010,zhao_symmetry_2010,bao_evidence_2012,maher_evidence_2013,velasco_jr_competing_2014,lee_chemical_2014,shi_energy_2016} and fractional\cite{kou_electron-hole_2014,maher_tunable_2014,shi_energy_2016} filling.
However, these experiments are insufficient to constrain realistic theoretical models, in which the preferred ordering is determined by a combination of
the Zeeman energy, which splits the spins; Coulomb interactions and band structure effects, both of which distinguish between the $N=0, 1$ orbitals; and several small ``valley anisotropies'' which weakly break the valley-SU(2) symmetry\cite{barlas_intra-landau-level_2008, abanin_charge_2009, cote_orbital_2010, jung_lattice_2011, kharitonov_canted_2012,shizuya_structure_2012, lambert_quantum_2013}.  Indeed, two recent experimental papers explain their data using mutually contradictory single-particle\cite{lee_chemical_2014} and purely interacting pictures\cite{kou_electron-hole_2014}.

Constructing a more complete theory of symmetry breaking in bilayer graphene requires experimental determination of the partial filling of each spin, valley, and orbital level, $\nu_{\xi N\sigma } = \langle \hat{N}^e_{\xi N\sigma } \rangle / N^\Phi$ as it evolves with total LL filling.
Here we introduce a direct measurement of two out of three of these components, by
 exploiting the fact that the four valley and orbital components indexed by $\xi N$ have different weights on the two layers of the bilayer.  We detect this  difference in layer polarization capacitively, and use it to infer the fillings $\nu_{\xi N}$ as a function of both the total electron density and applied perpendicular electric field.

	\begin{figure*}[ht!]
	\begin{center}
\includegraphics[width=140 mm]{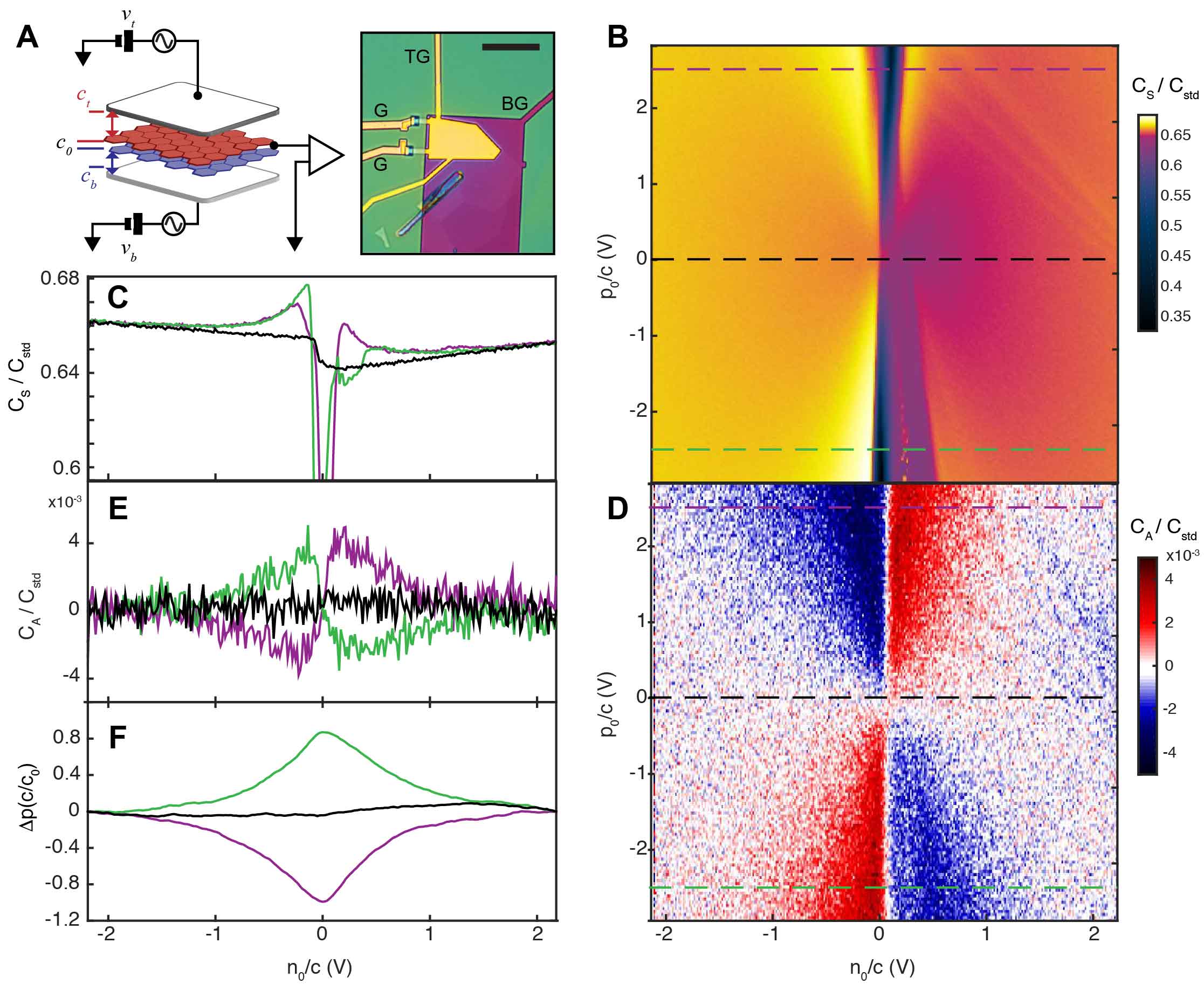}
\caption{
\textbf{Layer polarization at zero magnetic field.} \textbf{(A)} Left: Measurement schematic showing geometric gate capacitances $c_t$ and $c_b$ and interlayer capacitance $c_0$. Capacitance is measured using a cryogenic bridge circuit by comparison with a standard capacitor $C_{std}$, measured to be 404$\pm$20 fF (see SI). Right:  Device image. Top gate (TG), back gate (BG) and contacts to bilayer graphene (G) are shown.  Scale bar is 10 $\mu$m; device area is approximately 87 $\mu$m$^2$.
\textbf{(B)} $C_S$ measured at $B=0$ and $T=1.6$K as a function of $n_0/c = v_t + v_b$ and $p_0/c = v_t - v_b$. A $p_0$-dependent band gap is visible as the dark region near $n_0=0$.
\textbf{(C)} Line traces taken at different values of $p_0$, corresponding to dashed lines in (B).  Band edge van Hove singularities\cite{young_capacitance_2011} and electron-hole asymmetry\cite{henriksen_measurement_2010} are both evident.
\textbf{(D)} $C_A$ measured under the same conditions.  A common, constant background has been subtracted to account for fixed parasitic capacitances.
\textbf{(E)} Line traces at different values of $p_0$ corresponding to dashed lines in (D).
\textbf{(F)} Integrated change in polarization, $\frac{c_0}{c}\int C_A \, d(\frac{n_0}{c})=\Delta p$, with the constant of integration fixed to be zero at high $|n_0|$.  In accordance with single particle band structure\cite{young_capacitance_2011},
wavefunctions are layer unpolarized for $p_0=0$, while for large $|p_0|$ the polarization peaks at $n_0=0$, where band wavefunctions are strongly layer polarized.}
		\label{fig1}
	\end{center}
\end{figure*}

	\begin{figure*}[ht!]
	\begin{center}
\includegraphics[width=160 mm]{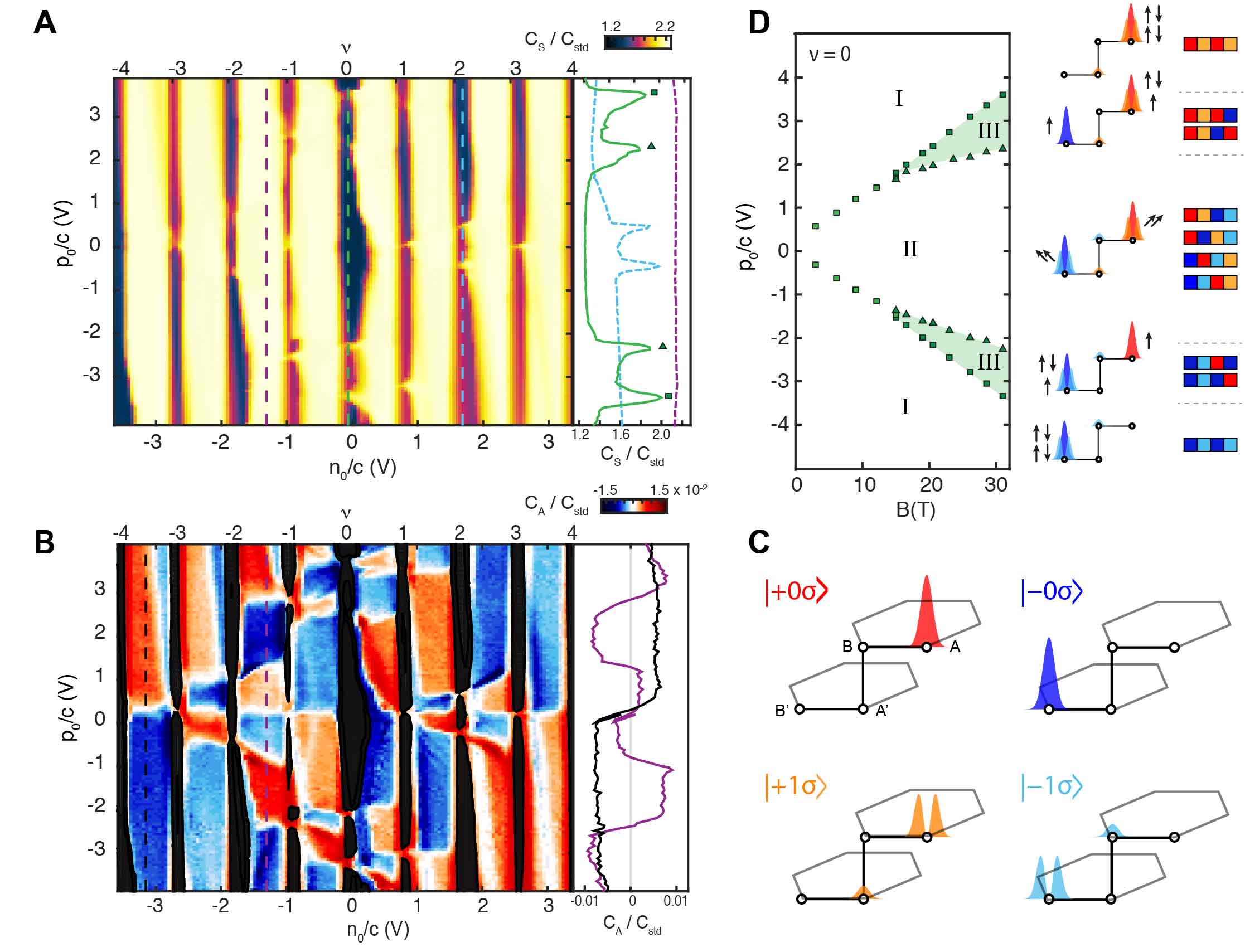}
		\caption{\textbf{Valley and orbital polarization of the zero Landau level.}
		\textbf{(A)} Layer-symmetric capacitance $C_S \propto \frac{\partial n}{\partial  n_0}$ at $T=300$~mK and $B=31$T. Incompressible states manifest as drops in $C_S$ (black) at all integer fillings $\nu$.  Phase transitions between different valley and orbital fillings at fixed $\nu$ manifest as compressible spikes, as shown in the side panel for $\nu=0$ (green, solid) and $\nu=2$ (light blue, dashed).  A total of 16 phase transitions are observed at integer $\nu$, with one each at $\nu=\pm3$, two at $\nu=\pm2$, three at $\nu=\pm1$, and four at $\nu=0$. No experimental contrast is visible at non-integer filling (purple, dashed). 	
		\textbf{(B)} Layer-antisymmetric capacitance $C_A \propto \frac{\partial p}{\partial  n_0}$ at $T=300$~mK and $B=31$T.  Black regions mask portions of the parameter space with large dissipation in $C_S$, which arises when a large gap  leads to a low in-plane conductivity and failure to charge regions of the sample during a $\sim$~13~$\mu$s measurement cycle (see SI). The color scheme highlights the 4-tone contrast, interpreted as filling of $|\xi N \sigma\rangle$ =  $|+0\sigma\rangle$ (red), $|+1\sigma\rangle$ (orange), $|-0\sigma\rangle$ (blue), and $|-1\sigma\rangle$ (cyan) LLs.   
		\textbf{(C)} Schematic depiction of the four single-particle wavefunctions $| \xi N\sigma\rangle$, showing their relative support on the four atomic sites $A$,$B$,$A'$ and $B'$ of the bilayer graphene unit cell. While the $|+0\sigma\rangle$ levels are fully polarized ($\alpha_0=1$), we calculate $\alpha_1=0.63$ for the $|+1\sigma\rangle$.
		\textbf{(D)} Left: phase diagram of gapped states at $\nu=0$.  Points are experimentally determined by measuring peaks in $C_S$, as in (A) (green dashed line), for $0<B<31$ T.  At high $B\gtrsim15$T an intermediate phase III emerges between the layer-unpolarized canted antiferromagnetic phase II and the layer-polarized phase I \cite{weitz_broken-symmetry_2010, maher_evidence_2013}.  Center: schematics of the layer, orbital and spin polarizations of phases I, II, and III.  Right: ten distinct filling sequences that determine the three valley and orbital polarizations of phases I, II and III. These sequences are extracted from Fig. 2b, filling from $\nu=-4$ to $\nu=0$ over the full range $-4V<p_0/c<4V$.}
	\label{fig2}
	\end{center}
\end{figure*}

	\begin{figure*}[ht!]
	\begin{center}
\includegraphics[width=160 mm]{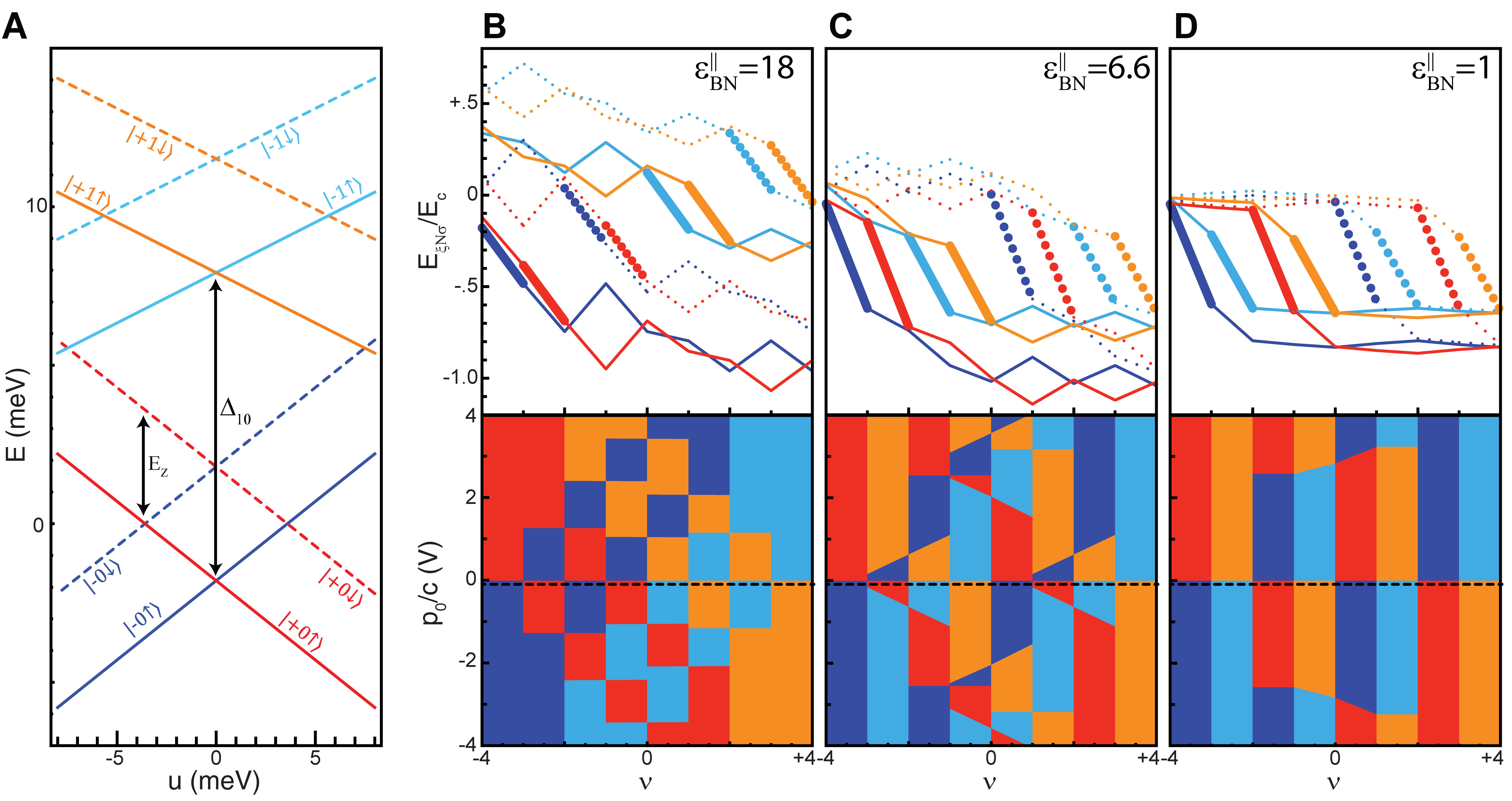}
		\caption{\textbf{Theoretical model of spin, valley, and orbital anisotropies in the zero Landau level}
\textbf{(A)} Single particle energy spectrum of the ZLL at $B=31$T derived from a four-band tight binding model\cite{jung_accurate_2014} (see also Eq. \ref{eq:free}).  The interlayer potential $u$ couples to the layer polarization of each state as $\xi\alpha_N u$, differing in sign for the two valleys and magnitude for the two orbitals; the spin degeneracy is lifted by the Zeeman energy $E_Z \approx$3.6~meV; and  the $N=0$ and $1$ orbitals are split by the band structure paramter $\Delta_{10} \approx$ 9.7~meV.
\textbf{(B-D)}
\textit{Bottom panels}:  Hartree-Fock phase diagram for three regimes: (\textbf{B}) negligible Coulomb interactions, $\epsilon_{\textrm{BN}}^\parallel=18$, (\textbf{C}) best fit of the theoretical model, $\epsilon_{\textrm{BN}}^\parallel=6.6$, and (\textbf{D}) strong interactions, $\epsilon_{\textrm{BN}}^\parallel=1$.
Colors blue, cyan, red and orange  indicate whether levels of type $\xi N = -0, -1, +0, +1$ are filling, so that the result should mimic the  observed $C_A$. Case (C) shows good agreement with the experimental $C_A$-data  shown in  Fig. 2b, while interactions which are too weak (B) or strong (D) do not reproduce the observed filling sequences.
The black dashed line indicates a cut at $p_0/c=-100$meV  corresponding to the top panels.
\textit{Top panels}: level filling schematic for $\frac{p_0}{c} = -100$mV.
Within the Hartree-Fock  approximation, we calculate the energy to add an additional electron to  level $\sigma \xi N$ given the current filling $\{ \nu_{\sigma \xi N} \}$, generating eight curves $\epsilon_{\sigma \xi N}(\nu)$ which change with the total filling $\nu$.
Colors indicate the $\xi N$ index of the level, while solid vs. dashed indicates the spin. The bold portion indicates the range of $\nu$ over which the level is coincident with the Fermi energy.
As isospin $\sigma \xi$ fills, both of its $N=0$ and $1$ orbitals decrease in energy due to favourable Coulomb correlations,  while the components of the opposite valley (i.e., layer)  decrease slightly in energy due to the capacitance of the bilayer.	
The relative magnitude of these effects, combined with the single-particle splittings, determines the filling order for the three interactions strengths (B), (C), (D).
\label{fig3}}
	\end{center}
\end{figure*}

	\begin{figure}[t!]
	\begin{center}
\includegraphics[width=89 mm]{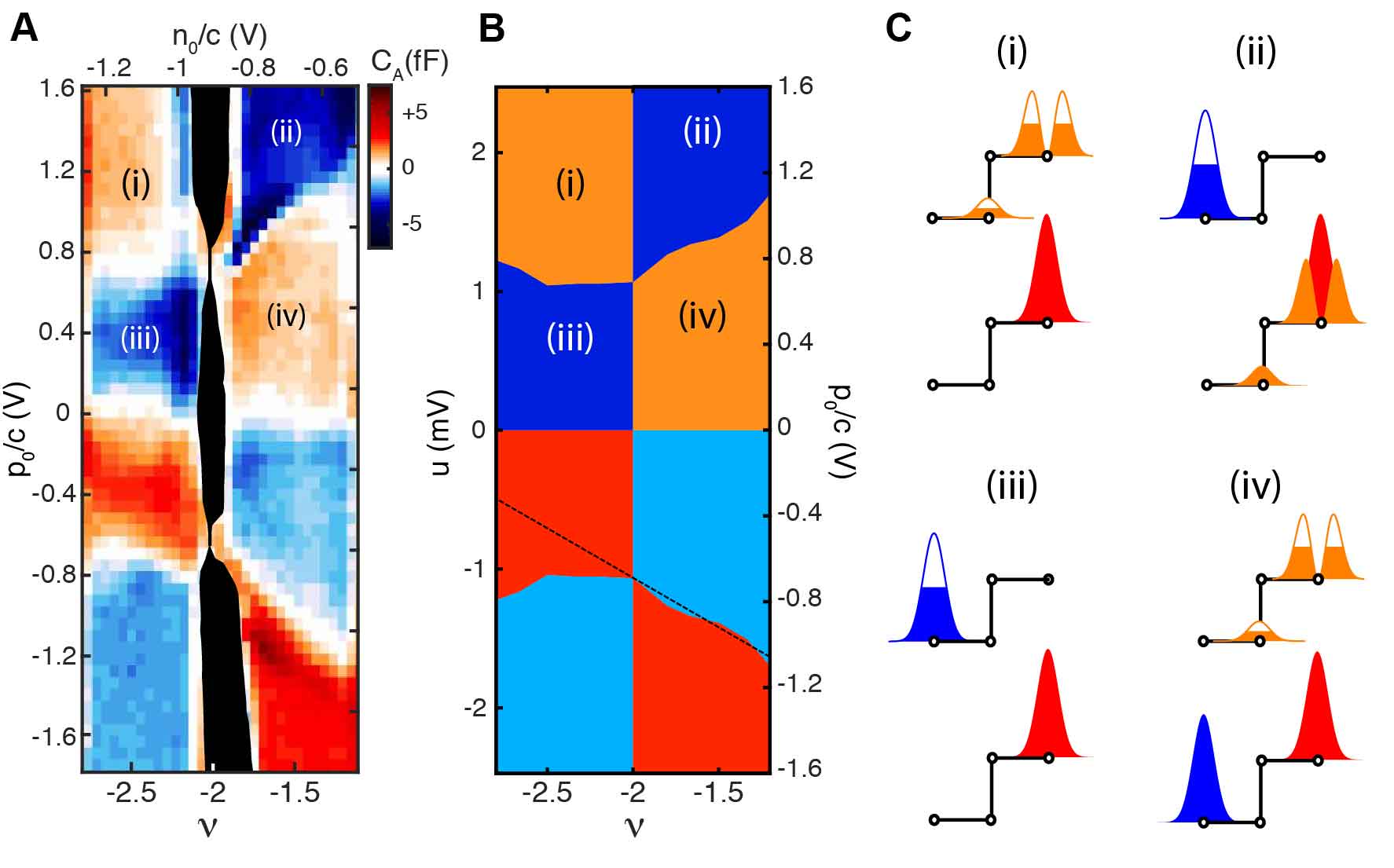}
		\caption{\textbf{Correlation effects at partial Landau level filling. } \textbf{(A)} Detail of $C_A$ near $\nu=-2$. Sign changes in $C_A$ as a function of $n_0$ indicate polarization extrema and phase transitions.   The strong dependence of the phase transition lines on $\nu$ are a consequence of interactions, which cause the energy per particle of the competing phases to depend on $\nu$.
\textbf{(B)} Phase diagram from multicomponent infinite DMRG calculations\cite{zaletel_infinite_2015} for a four-band model of bilayer graphene with Coulomb interactions.
In contrast to the Hartree-Fock prediction shown in Fig. 3C, experiments show that the magnitude of the slope of the  phase boundary between (i) and (iii)  differs from the boundary betweem (ii) and (iv).
This implies that strong scattering between the $N=0, 1$ orbitals breaks the particle-hole symmetry $\nu \leftrightarrow -(\nu + 2)$,  an effect which is correctly accounted for in our DMRG simulations.
\textbf{(C)} Schematic representation of the four phases appearing in (A).  Each of the four  orbital types $\xi N$ depicted in Fig. 2C can either be filled (solid, shown in bottom unit cell), or in the process of filling (partially solid, shown in top unit cell).
Phase (ii), for example, consists of fully filled $|+0\rangle$ and $|+1\rangle$ levels, while level $|-0\rangle$ is filling.
\label{fig4}}
	\end{center}
\end{figure}

Our devices consist of hexagonal boron nitride encapsulated B-BLG flakes\cite{wang_one-dimensional_2013} fitted with metal top and bottom gates (Fig.~\ref{fig1}a).
The layer polarization and total charge density are tuned by a combination of applied  top and bottom gate voltages ($v_t$ and $v_b$), expressed through their symmetric and antisymmetric combinations $n_0(p_0)\equiv c_t v_t\pm c_b v_b$ with $c_{t(b)}$ the geometric capacitances between the respective gates and the B-BLG.  $-n_0$ and $-p_0$ correspond to the induced charge density and layer polarization in the limit of a perfectly metallic, infinitesimally spaced bilayer. Generically, the physically realized total density ($n$) and layer density imbalance ($p$) deviate from this limit, particularly at high magnetic fields.  A simple electrostatic model (see SI) shows that these deviations manifest as corrections to the measured gate capacitances ($C_T$ and $C_B$) as
\begin{align}
C_S\equiv C_T+C_B&=2c \frac{\partial n}{\partial n_0}\\
C_A \equiv C_T-C_B&=2c \frac{\partial n}{\partial p_0}=\frac{c^2}{c_0} \frac{\partial p}{\partial n_0},
\label{eq1}
\end{align}
where $c=(c_t+c_b)/2\approx$1.36 fF/$\mu$m$^2$ and $c_0$ is the interlayer capacitance of the bilayer. The antisymmetric combination $C_A$ is unique to bilayer systems, vanishing identically in a monolayer system, allowing us to measure the layer polarization $p$. Measuring $C_A$ is technically challenging, as it arises from series combination of the large interlayer capacitance and the comparatively small gate capacitance.  It is imperative that the dielectric layers be highly uniform across the device, that  the areal mismatch between top and bottom gate be kept $\ll 1\%$, and the geometric capacitance of the two gates be nearly the same; these requirements are difficult to satisfy in conventional two dimensional electron bilayers but are readily achieved using single crystal hBN gate dielectrics and atomically thin bilayers.

Figures\ref{fig1}b-c show $C_S$ and $C_A$ measured at zero magnetic field as a function of $n_0$ and $p_0$.
The $C_S$ data are dominated by quantum capacitance features of the B-BLG band structure, which features a quadratic band touching at low energies and hyperbolic bands at high energies\cite{mccann_electronic_2013}.  An electric field ($p_0$) opens a band gap with $\sqrt{\epsilon}$ van Hove singularities at the band edges, as can be readily seen in the experimental data (Fig.~\ref{fig1}c).
Although $C_S$ is approximately particle-hole symmetric, significant symmetry breaking contributions are evident. We attribute this to the skew interlayer hopping parameter $\gamma_4$ in the Slonczewski-Weiss-McClure model for graphite, which breaks particle-hole symmetry by making the lattice non-bipartite (see Supp. Fig. S2)\cite{henriksen_measurement_2010}.
$C_A$ data, in contrast, reflect the layer-resolved properties of the band wavefunctions (Fig.~\ref{fig1}d,e).
For $p_0\neq 0$, wavefunctions are layer-polarized near the band extrema, so that the first electrons or holes added to the neutral system are added to the corresponding low-energy layer. Reversing $p_0$ inverts the role of the top and bottom layers, inverting the sign of the measured signal with respect to $n_0$. At high overall electron density, the applied $p_0$ is fully screened, so that charge is added symmetrically to the two layers\cite{young_capacitance_2011}.  The relative layer polarization at different values of $n_0$ can then be extracted by integrating Eq.~\ref{eq1} with respect to $n_0$ (Fig.~\ref{fig1}f).

Fig.~\ref{fig2}a shows $C_S$ measured in the same device at $B=31$T in the ZLL.
We observe insulating states at all integer LL filling factors $\nu$, which are characterized by low capacitance $C_S$ and high dissipation (See Fig. S7 and S8 for dissipation data).
Adjusting $p_0 / c$ at fixed integer $\nu$ drives transitions characterized by a spike in $C_S$ indicating increased  conductivity\cite{maher_tunable_2014} and  compressibility, consistent with a closing of the charge gap.  16 such phase transitions are evident in the $C_S$ data.
Similar transitions have been reported in the literature: Ref. \onlinecite{weitz_broken-symmetry_2010} reported phase transitions at $\nu=\pm2$ and $p_0=0$, as well as a single phase transition at $\nu = 0$ and finite $p_0$. More recently, the $p_0=0$ transitions at $\nu=\pm1,\pm3$ are evident in Ref.~\onlinecite{maher_tunable_2014}, while the splitting of the $p_0=0$ phase transition at $\nu=\pm2$ suggests the formation of a stable, gapped, layer unpolarized state in the region $p_0 \sim 0$, as was reported in Refs. \onlinecite{maher_tunable_2014,velasco_jr_competing_2014}. Only in Ref.~\onlinecite{lee_chemical_2014} was a potential gapped phase observed at intermediate $p_0$ and $\nu=0$. However, a unified framework for understanding the diverse competing phases has not yet emerged.

In contrast to layer-insensitive capacitance\cite{feldman_broken-symmetry_2009,kou_electron-hole_2014,lee_chemical_2014} and transport\cite{zhao_symmetry_2010,weitz_broken-symmetry_2010,maher_evidence_2013,velasco_jr_competing_2014,maher_tunable_2014} measurements, $C_A$ provides high experimental contrast throughout the $n_0$-$p_0$ plane (Fig.~\ref{fig2}b).
Strikingly, the measured $C_A$ falls into discrete levels, corresponding to blue, cyan, orange, and red on the color scale of Fig. \ref{fig2}b. The transitions observed in $C_S$ at integer filling fall on lines in the $n_0-p_0$ plane along which the sign of $C_A$ changes abruptly.
To understand the color scale quantitatively, we compute the layer polarization of the ZLL single-particle eigenstates.
B-BLG has a 4-site unit cell, with sites $A, B$ in the top layer and $A', B'$ in the bottom (see Fig. 2c), and hence the ZLL wavefunctions decompose into their components on the four sublattices, $\mathbf{\Psi}(x) = \left(\phi_A(x),\phi_{B'}(x),\phi_{B}(x),\phi_{A'}(x) \right)$.
The layer polarization $\alpha_{\xi N \sigma}\equiv \int d^2x |\phi_A(x)^2| + |\phi_{B}(x)^2| - |\phi_{A'}(x)^2| - |\phi_{B'}(x)^2|$ is constant across all states in a LL, and independent of spin.
It has opposite sign in the two valleys\cite{mccann_electronic_2013}, so that positive and negative $C_A$ correspond to filling valley K and K', and its magnitude depends on the orbital quantum number, so that $\alpha_{\xi N \sigma} = \xi \alpha_N$. At $B=31$T, band structure calculations\cite{jung_accurate_2014} show that $\alpha_0 = 1$ and $\alpha_1 = 0.63$.

As electrons enter LL $|\xi N \sigma \rangle$ they contribute a polarization whose magnitude and sign depend on the
level being filled. Since $C_A \propto \frac{\partial p}{\partial n_0}$, where $n_0$ is very nearly the electron density, we thus interpret red, orange, blue and cyan as indicating regions where electrons are filling $|\xi N\rangle$ = $|+0\rangle$, $|+1\rangle$, $|-0\rangle$, and  $|-1\rangle$ type LLs respectively. This supports a scenario in which, away from phase boundaries, only one of these LLs is filling at each particular $(n_0, p_0)$. Indeed, numerical calculations (see Supp. Fig. S5) show that as isospin $\xi \sigma$ fills, around 90\% of the electrons enter into either the $N=0$ or $N=1$ orbital; e.g., either $\frac{\partial \nu_{\xi 0 \sigma }}{\partial \nu} \gsim 0.9$ or $\frac{\partial \nu_{\xi 1\sigma }}{\partial \nu} \gsim 0.9$, according to whether the region is red/blue or orange/cyan respectively.

The polarization of all gapped integer states can now be obtained by summing the level filling sequence starting from the $\nu=-4$ vacuum.
Consider $\nu = 0$, where five incompressible states are visible in $C_S$ (Fig. \ref{fig2}d).
The order in phase I, at large $p_0 > 0$, can be inferred from the observed $C_A$ sequence of red, orange, red, orange, implying two $N=0$ and two $N=1$ states are filled in valley $\xi = +$.  Its layer-inverse occurs in valley $\xi  = -$ for large $p_0 < 0$.  The phase I states are fully valley polarized, and hence spin  and orbitally unpolarized due to Pauli exclusion.
Phase II, at $p_0$ near zero, fills levels $|+0\rangle, |+1\rangle, |-0\rangle, |-1\rangle$.
A state which fills both orbitals of opposite valleys is consistent with the canted anti-ferromagnetic state\cite{kharitonov_canted_2012,maher_evidence_2013}; however, from $C_A$ alone we cannot infer whether the spins are polarized or canted.
Finally, phase III at intermediate $|p_0|$ fills levels $\pm0, \pm1, \pm0, \mp 0$.  

The orbital and valley filling sequences derived from Fig.~2B provide a more stringent set of constraints on theoretical models than the integer phase transitions alone.
For example, a single particle, four-band tight-binding model accounts for the correct number of integer $\nu$ transitions \cite{lee_chemical_2014} but fails when compared to the $C_A$ filling sequence. The single particle energies of the ZLL,
\begin{align}
E^{(1)}_{\xi N\sigma } = -E_Z \sigma + N \Delta_{10} -\frac{u}{2} \xi \alpha_N.
\label{eq:free}
\end{align}
are shown in Fig. 3A for B=31T.  Here $E_Z$ is the Zeeman energy, $u$ is the potential across the bilayer, and $\Delta_{10} \propto \gamma_4 B$ is the splitting between the $N=0, 1$ orbitals which arises because particle-hole symmetry no longer pins the ZLL energies to zero.
The interlayer bias $u\propto p_0$ couples to the differing layer polarization of each state, $\xi\alpha_N$, leading to $u$-tuned crossings at integer fillings; however, the single particle picture predicts that $N=0$ levels fill first for all but the very highest $u$ (Fig.~3B), and the transitions would have zero slope in the $n_0 - p_0$ plane.
This disagrees with the experimental  $C_A$ data where $N=0$ and $N=1$ states of the same valley often fill in sequence, and several transition lines have a significant slope.

The departure from the single-particle picture arises from the failure to account for the Coulomb interaction, which both favors sequential filling of $N=0,1$ levels of the same isospin due to exchange\cite{abanin_charge_2009,barlas_intra-landau-level_2008} and penalizes valley polarization due to the interlayer capacitance of the bilayer. However, if Coulomb correlations are made too strong, the first effect dominates and $N=0, 1$ orbitals of an isospin always fill sequentially, as shown in Fig.~3D. Apparently  competition between the splitting $\Delta_{10}$ and Coulomb correlations is essential.

To determine whether a single model can account for all the observed phase boundaries, we analyze a model which accounts for the single-particle splittings, the  SU(4)-invariant screened Coulomb interaction, and several subleading ``valley-anisotropies,'' as detailed in the supplementary materials.  Evaluating the model within a Hartree-Fock approximation allows us to compute the energies of the competing filling sequences, and thereby determine their phase boundaries in the $n_0 - p_0$ plane. The model depends on three phenomenological constants (a screening strength, the perpendicular dielectric constant of the BLG, and a valley anisotropy),  which we can now estimate by matching the location of the integer transitions and their dependence on an in-plane $B$-field. The resulting phase diagram is shown in Fig. 3C, and shows good agreement with the $C_A$-data of Fig. 2B, including the location of the transitions in absolute units of $p_0/c$ and several of their slopes.
In our model, each integer state is obtained by fulling filling
some number of $\xi N \sigma $-levels; in particular it does not require ``interlayer coherent phases'' which spontaneously break the valley U(1)$\times$U(1) symmetry, in contrast to  theories which predict such phases at $\nu = \pm 1, p_0 =0$.
We also predict that the $\nu=0$ phase III observed at $\nu=0$ likely hosts helical edge states similar to those recently described in twisted bilayer graphene\cite{sanchez-yamagishi_observation_2016}. This state is stabilized by the single  particle anisotropy $\Delta_{10}$ and antagonized by the Coulomb interactions, suggesting it could be further stabilized in devices with stronger screening due  to proximal metal gates.

Despite good overall agreement, there is an interesting qualitative discrepancy between the Hartree-Fock analysis and the data.
In the experiment, the slope of the phase transition line between $-2 < \nu < -1$ (see Fig. 4a, boundary (ii) - (iv) ) is significantly larger than the slope of the adjoining transition across $-3 < \nu < -2$ (boundary (i)-(iii) of Fig. 4a).
Within the Hartree-Fock approximation, the slopes are identical, and in fact any model  which  neglects scattering (e.g., ``Landau-level mixing'') between  the $N=0, 1$ orbitals has a particle-hole symmetry $\nu + 2  \to - (\nu + 2), u - u_\ast(\nu = -2) \to -(u - u_\ast(-2))$, forcing  the two boundaries to mirror  each other.
To account for the asymmetry, we instead find the ground state of the model's Hamiltonian using the  multicomponent infinite-density matrix renormalization group, which takes full account of correlations. \cite{zaletel_infinite_2015}
Allowing LL mixing produces a kink in the slope at $\nu=-2$, as experimentally observed (Fig. 4b.).  LL mixing is known to generate effective three-body interactions which stabilize fractional non-Abelian phases\cite{moore_nonabelions_1991, Rezayi_breaking_2011}; our results suggest that these interactions may be  stronger in BLG than in conventional semiconductor quantum wells.

In conclusion, we have described a new experimental technique to determine the layer polarization of van der Waals bilayers and used it to constrain a detailed model of symmetry breaking in the bilayer graphene ZLL. Our technique is readily applicable to quantitatively probe layer, valley, and spin polarization in other atomic layered materials, including twisted bilayer graphene and both homo- and heterobilayers of transition metal dichalcogenides.

\vspace{5mm}

\paragraph*{Contributions}
AFY designed the experiment.  LW, JIAL, and AAZ fabricated devices, supervised by AFY, CRD and JH. AFY, BMH, and AAZ acquired and analyzed experimental data.  MZ and AFY built and analyzed the theoretical model. TT and KW grew the hBN crystals.  RCA, BMH, AFY, and MZ wrote the paper, and all authors commented on the the manuscript.

\paragraph*{Acknowledgements}
We acknowledge helpful discussions with E. Berg, L. Levitov, A. Macdonald, and I. Sodemann. The work at UC Santa Barbara was funded by an NSF EAGER under DMR-1636607.  The work at MIT was funded by the BES Program of the Office of Science of the US DOE, contract no. FG02-08ER46514, and the Gordon and Betty Moore Foundation, through grant GBMF2931.  Work at Columbia was funded by the NSF MRSEC under DMR-1420634, and by the US ONR under N00014-13-1-0662.



\section*{References}
\renewcommand\bibsection{%
  \paragraph*{}%
}%


\setcounter{figure}{0}
\setcounter{section}{1}
\setcounter{subsection}{0}

\renewcommand\thesection{S\arabic{section}}
 \renewcommand{\thefigure}{S\arabic{figure}}
\renewcommand{\bibnumfmt}[1]{[S#1]}
\renewcommand{\citenumfont}[1]{S#1}
\newpage

\section*{Supplementary Information}
\tableofcontents
\setcounter{section}{0}
\section{Experimental methods}
\subsection{Sample preparation \label{sec:sampleprep}}
Bilayer graphene samples encapsulated in hexagonal boron nitride were fabricated using a dry transfer method\cite{Swang_one-dimensional_2013}.  Particular care is taken to ensure that the top and bottom hBN flakes are of the same thickness, measured by atomic force microscopy to be 19 and 20 nm, respectively.  During fabrication, care is also taken to minimize the area of graphene bilayer gated by only one of the two gates, as single-gated areas contribute a systematic error to the measured $C_A$ signal proportional to the area and to $C_S$.  Anticipating $\frac{C_A}{C_S}\approx \frac{c}{2c_0}\lesssim \frac{3.35 \textrm{\textrm{\AA}}}{39\mathrm{nm}}=.0086$ , we ensure that the areal mismatch between bottom- and top-gated areas is less than .5\%.
\subsection{Capacitance measurement electrical schematic}

\begin{figure}[b!]
	\begin{center}
\includegraphics[width=\columnwidth]{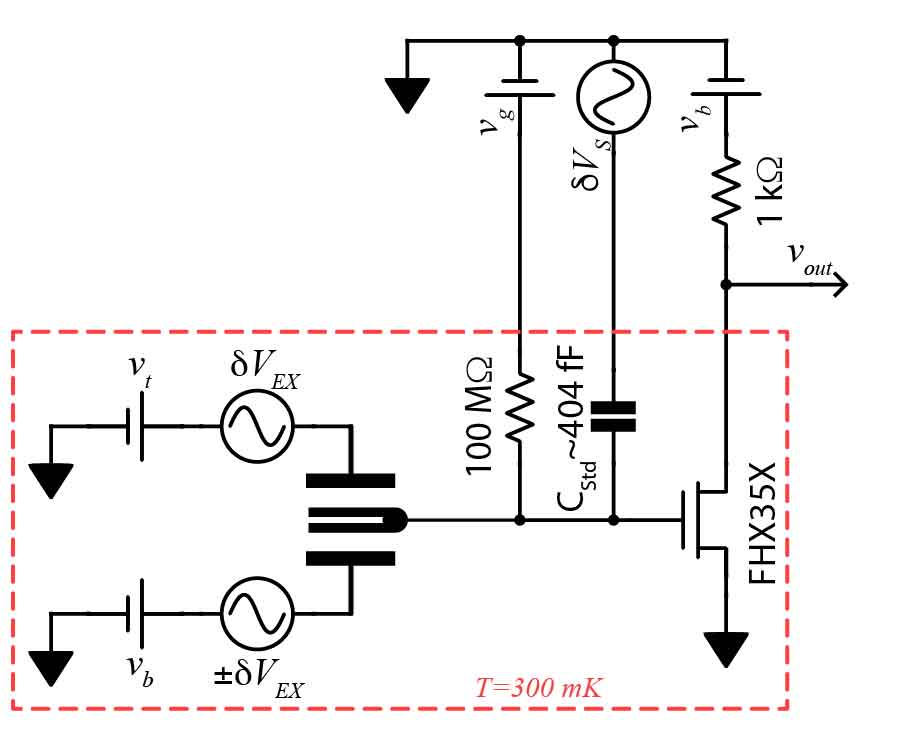}
		\caption{Electrical schematic of the capacitance measurement. }
		\label{fig:electrical}
	\end{center}
 \end{figure}

 Capacitance measurements were made using a cryogenic impedance transformer based on an FHX35X high electron mobility transistor\cite{Sashoori_single-electron_1992} in a bridge configuration connected to the bilayer graphene ohmic contacts (see Fig. \ref{fig:electrical}). $v_g$ sets the transistor bias point and $v_b$ adjusted to be sufficiently low that no sample heating is observed. To measure $C_{S(A)}$, two synchronized and nearly equal-magnitude AC signals ($\delta V_{EX}$)are applied to the top and bottom gates, whose relative magnitude is chosen to match the ratio of geometric capacitances $c_t/c_b$ extracted from the DC characteristics of the device. The signals are applied in phase for $C_S$ and out of phase for $C_A$.  A third AC signal is applied to a standard capacitor $C_{std}$ with amplitude and phase that null the signal at the input of the cryogenic amplifier, and the capacitance and dissipation determined from the relation of the AC signals. $C_{std}$ was measured to be 404 fF during the cooldown in which the data of Fig. 1 of the main text were measured.  We used this value to determine $C_S$ and $C_A$ shown in subsequent figures, although in our experience $C_{std}$ can vary by up to 20 fF from cooldown to cooldown, thus introducing a systematic uncertainty of approximately 5\% in the capacitance shown in Figs. 2, 3 and 4. All data shown are acquired off-balance, by monitoring the voltage at the balance point as DC values of the gate voltages are changed.  Data in Fig 2a and 2b are measured at 67.778 kHz using a 10 mV variation of $n_0/c$ and of $p_0/c$, respectively.

Interpretation of $C_A$ as a thermodynamic derivative requires that the sample is sufficiently conductive to fully charge over a time scale comparable to the inverse measurement frequency\cite{Sgoodall_capacitance_1985}.  At low temperature and high magnetic fields, our sample becomes strongly insulating at integer filling factors, precluding this condition being satisfied. Failure to charge manifests as an increase in the out of phase, dissipative signal in the capacitance, allowing us to monitor charging across the parameter range.  In Figs. 2B and 4A, regions in which the sample does not charge are masked in black, and dissipation data for all data sets is shown in Figs. \ref{CapAndDis1}-\ref{CapAndDis2}.

\subsection{Electrostatic model of bilayer graphene capacitance measurements}
We model the dual gated graphene bilayer as a four plate capacitor, with the $c_i$ corresponding to the geometric capacitances as indicated Fig. 1B (inset).  The $n_i$ denote the areal charge carrier densities on the four plates.  Equations for the charge stored on each capacitor plate, as well as overall charge neutrality, result in four equations,
\begin{align}
  n_t+n_1+n_2+n_b&=0 \label{elec1}\\
  c_t(v_t-\phi_1)&=n_t  \label{elec2}\\
  c_0(\phi_1-\phi_2)&=\frac{n_t+n_1-n_2-n_b}{2}  \label{elec3}\\
  c_b(\phi_2-v_b)&=-n_b.
  \label{elec4}
\end{align}
Note that in the language of the main text, $\phi_2-\phi_1=u$.  Eqs. \ref{elec1}-\ref{elec4} are supplemented by the condition of electrochemical equilibrium between the top and bottom layers of the bilayer,
\begin{align}
\phi_1&=v_0-\mu_1\label{elec5}\\
\phi_2&=v_0-\mu_2\label{elec6}
\end{align}
where $v_0$ is the voltage applied to the bilayer, and $\mu_i$ is the chemical potential on layer $i$.  The $\mu_i$ depend on both $n_1$ and $n_2$ through the constitutive relations that derive from the electronic structure of the bilayer.

Capacitance measurements are performed with a small AC signal applied to one of three terminals while the corresponding variation in charge density is read out on another terminal.  For small variations, then, the differential versions of Eqs. \ref{elec1}-\ref{elec6} are relevant.  In particular, Eqs. \ref{elec5},\ref{elec6} can be expressed in terms of the inverse compressibility matrix of the bilayer itself, $\kappa_{ij}=\partial \mu_i/\partial n_j$,
\begin{align}
\delta\phi_1&=\delta v_0-\kappa_{21}\delta n_1-\kappa_{22}\delta n_2\label{vars1}\\
\delta\phi_2&=\delta v_0-\kappa_{11}\delta n_1-\kappa_{12}\delta n_2
\label{vars2}
\end{align}
Note that $\kappa_{12}=\kappa_{21}$ follows from a Maxwell relation.

Experimentally, we measure the elements of the capacitance matrix
\begin{equation}
  C_{ij}\left(\{c\}, \{\kappa\}\right)=\left.\frac{\delta n_i}{\delta v_j}\right|_{\delta v_{k\neq j}=0}.
  \label{capmat}
\end{equation}
Where the indices indicate the voltages applied to the top gate ($v_t$), bottom gate ($v_b$), or bilayer itself ($v_0$).
Three elements of the capacitance matrix are independent, and we choose the most directly experimentally relevant combinations: the penetration field capacitance $C_P\equiv-C_{BT}=-C_{TB}$, and top and bottom gate capacitances $C_B\equiv C_{B0}$ and $C_T\equiv C_{T0}$.  Expressions for these three quantities can be found by varying Eqs. \ref{elec1}-\ref{elec4} and using Eqs. \ref{vars1}-\ref{vars2} to eliminate $\phi_1$ and $\phi_2$.  All measurable capacitances depend on all three components of the compressibility matrix,
\begin{widetext}
\begin{align}
C_P&=\frac{c_b c_t (c_0 \kappa_{11} \kappa_{22}+\kappa_{12}-c_0\kappa_{12}^2)}
{1 + (c_0 + c_b) \kappa_{11} + (c_0 + c_t)\kappa_{22} + (c_b c_t+ c_0 c_b + c_0 c_t)( \kappa_{11} \kappa_{22}-\kappa_{12}^2) -2 c_0 \kappa_{12}}\label{complic1}\\
C_B&=\frac{c_b\left(1+c_t(\kappa_{11}-\kappa_12)+c_0(\kappa_{11}+\kappa_{22}-2\kappa_{12})\right)}
{1 + (c_0 + c_b) \kappa_{11} + (c_0 + c_t)\kappa_{22} + (c_b c_t+ c_0 c_b + c_0 c_t)( \kappa_{11} \kappa_{22}-\kappa_{12}^2) -2 c_0 \kappa_{12}}\label{complic2}\\
C_T&=\frac{c_t\left(1+c_b(\kappa_{22}-\kappa_12)+c_0(\kappa_{11}+\kappa_{22}-2\kappa_{12})\right)}
{1 + (c_0 + c_b) \kappa_{11} + (c_0 + c_t)\kappa_{22} + (c_b c_t+ c_0 c_b + c_0 c_t)( \kappa_{11} \kappa_{22}-\kappa_{12}^2) -2 c_0 \kappa_{12}}.\label{complic3}
\end{align}
\end{widetext}

As described in the main text, the ultimate quantities of interest are the total density and layer density imbalance of the bilayer, $n\equiv n_1+n_2$ and $p=n_1-n_2$, while the most natural control parameters are $n_0=c_t v_t+c_bv_b$ and $p_0=c_tv_t-c_bv_b$.  The partial derivatives of $n$ with respect to $n_0$ and $p_0$ follow trivially from the fact that partial derivatives of the $n$ with respect to the gate voltages can be measured directly:
\begin{widetext}
\begin{align}
\frac{\partial n}{\partial n_0}
&=\frac{\partial n}{\partial v_t}\frac{\partial v_t}{\partial n_0}+\frac{\partial n}{\partial v_b}\frac{\partial v_b}{\partial n_0}
=\frac{C_T}{c_t}+\frac{C_B}{c_b}
=\frac{C_S}{c}\\
\frac{\partial n}{\partial p_0}
&=\frac{\partial n}{\partial v_t}\frac{\partial v_t}{\partial p_0}+\frac{\partial n}{\partial v_b}\frac{\partial v_b}{\partial p_0}
=\frac{C_T}{c_t}-\frac{C_B}{c_b}
=\frac{C_A}{c}.
\end{align}
\end{widetext}
For convenience we we have introduced the average geometric capacitance $c=\frac{c_b+c_t}{2}$ and the geometric capacitance asymmetry between top and bottom gates $\delta=\frac{c_b-c_t}{c_b+c_t}$; we also define
here the capacitance observables measured and described in the main text as

\begin{equation}
  C_{S(A)}\equiv \frac{C_B}{1-\delta}\pm\frac{C_B}{1+\delta}.
\end{equation}

Derivatives of $p$ can also be computed by varying \ref{elec1}-\ref{elec4} and using Eqs. \ref{vars1}-\ref{vars2}, to give expressions in terms of the $\kappa_{ij}$.  After solving Eqs. \ref{complic1}-\ref{complic3} for the three $\kappa_{ij}$ and substituting the results, we arrive at expressions for these derivatives in terms of observable capacitances ($C_P$, $C_S$, and $C_A$):

\begin{widetext}
\begin{align}
\frac{\partial p}{\partial n_0}&=
\frac{\partial n_1}{\partial v_t}\frac{\partial v_t}{\partial n_0}+\frac{\partial n_1}{\partial v_b}\frac{\partial v_b}{\partial n_0}-\frac{\partial n_2}{\partial v_t}\frac{\partial v_t}{\partial n_0}-\frac{\partial n}{\partial v_b}\frac{\partial v_b}{\partial n_0}
=\frac{\partial n_1}{\partial v_t}\frac{1}{c_t}+\frac{\partial n_1}{\partial v_b}\frac{1}{c_b}-\frac{\partial n_2}{\partial v_t}\frac{1}{c_t}-\frac{\partial n}{\partial v_b}\frac{1}{v_b}\\
&=\frac{c_t (\kappa_{11} + \kappa_{12}) + 2 c_0 (\kappa_{11} - \kappa_{22})-c_b (\kappa_{12} + \kappa_{22})}
{1 + (c_0 + c_b) \kappa_{11} + (c_0 + c_t)\kappa_{22} + (c_b c_t+ c_0 c_b + c_0 c_t)( \kappa_{11} \kappa_{22}-\kappa_{12}^2) -2 c_0 \kappa_{12}}\\
&= \frac{2c_0}{c(1-\delta^2)}\frac{C_S-4C_P-2c-\delta ((2 c + C_S) \delta -C_A (1 - \delta^2))}{c(1-\delta^2)}+\frac{4 C_P -C_S (1 - \delta^2))}{c(1-\delta^2)}\\
\frac{\partial p}{\partial p_0}&=
\frac{\partial n_1}{\partial v_t}\frac{\partial v_t}{\partial n_0}-\frac{\partial n_1}{\partial v_b}\frac{\partial v_b}{\partial n_0}-\frac{\partial n_2}{\partial v_t}\frac{\partial v_t}{\partial n_0}+\frac{\partial n}{\partial v_b}\frac{\partial v_b}{\partial n_0}
=\frac{\partial n_1}{\partial v_t}\frac{1}{c_t}+\frac{\partial n_1}{\partial v_b}\frac{1}{c_b}-\frac{\partial n_2}{\partial v_t}\frac{1}{c_t}-\frac{\partial n}{\partial v_b}\frac{1}{v_b}\\
&=\frac{2 + c_t (\kappa_{11} + \kappa_{12}) + c_b (\kappa_{12} + \kappa_{22})}
{1 + (c_0 + c_b) \kappa_{11} + (c_0 + c_t)\kappa_{22} + (c_b c_t+ c_0 c_b + c_0 c_t)( \kappa_{11} \kappa_{22}-\kappa_{12}^2) -2 c_0 \kappa_{12}}\\
&=\frac{2c_0}{c(1-\delta^2)}\frac{C_A(1-\delta^2)+\delta(CS(1-\delta^2)-4 CP + 2 c (1 - \delta^2))}{c(1-\delta^2)}+\frac{4 C_P \delta -C_A (1 - \delta^2)}{c(1-\delta^2)}
\end{align}
\end{widetext}
%
%
\begin{figure*}[t]
	\begin{center}
\includegraphics[width=.4\columnwidth]{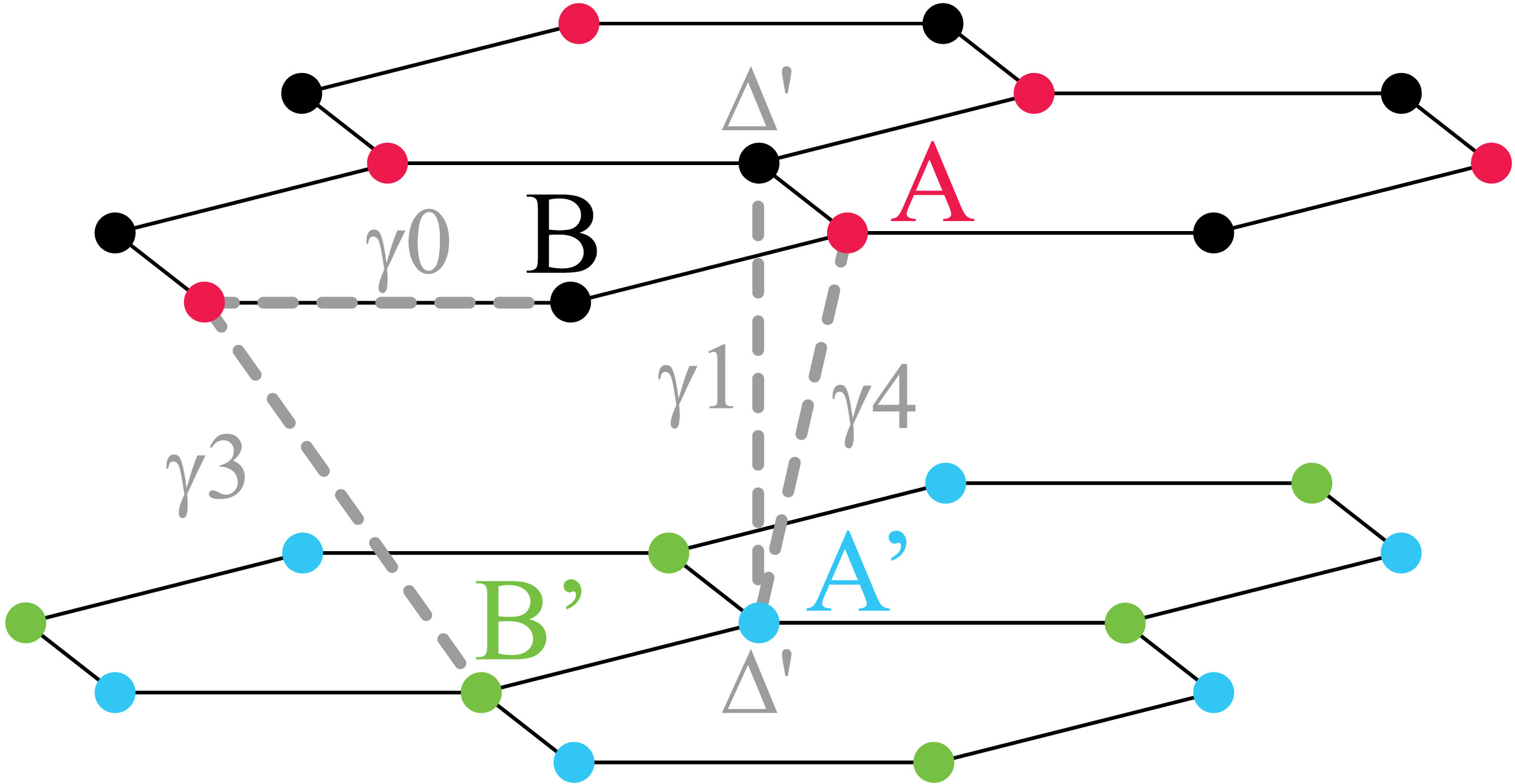}
		\caption{Bilayer graphene hopping parameters.}
		\label{fig:hopping}
	\end{center}
 \end{figure*}

In our device the top and bottom gate geometric capacitances are nearly symmetric with $\delta\approx .029$; in addition, we estimate that $\frac{c}{2c_0}\lesssim .0086$. Taking the leading order terms in these two small parameters, we finally arrive at simplified expressions, some of which are used in the main text.

\begin{align}
\frac{\partial n}{\partial n_0} &\approx \frac{C_S}{c}     \\
\frac{\partial n}{\partial p_0} &\approx \frac{C_A}{c}     \\
\frac{\partial p}{\partial p_0} &\approx \frac{2c_0}{c}\frac{C_S-4C_P+2c}{c}\\
\frac{\partial p}{\partial n_0} &\approx \frac{2c_0}{c}\frac{C_A}{c}
\end{align}
\section{Theoretical model of symmetry breaking in the BLG ZLL}
\subsection{Outline}
As described in the main text, calculating a phase diagram of bilayer graphene that correctly captures the experimentally observed effects requires considering a variety of both single particle and correlated electron effects.
In this section we describe a minimal model which accounts for these effects, and compute ground state energies using Hartree-Fock and DMRG in order to determine the $\nu$-dependence of the phase boundaries.
The presentation is organized as follows:
The Hamiltonian is a sum of the single-particle and two-body terms,
\begin{align}
H = H^{(1)} + H^{(2)}.
\end{align}
We first describe the terms we choose to include in $H$.  To make numerical progress, we then project the problem into the ZLL.
Finally, in order to compute the energies of the competing phases at arbitrary $\nu$, we perform both Hartree-Fock and DMRG computations.
\subsection{Single-particle Hamiltonian}.

We start with a tight binding model of bilayer graphene that includes three intersite hopping terms ($\gamma_0$, $\gamma_1$, and $\gamma_4$ as well as the dimer on-site energy ($\Delta'$) but neglect the trigonal warping term ($\gamma_3$). A diagram of the bilayer graphene structure with relevant hopping integrals is shown in Fig. \ref{fig:hopping}). At low energies, the Fermi surface has two disconnected parts near the corners of the Brillouin zone around the two inequivalent K points.  The single particle Hamiltonian can then be reduced to a four-band model corresponding to the four sites of the BLG unit cell, resulting (in, e.g., valley $K$) in
\begin{widetext}
  \begin{align}
\hat H_K^{B=0}  &=\left(
  \begin{array}{cccc}
  \frac{u}{2}           &0                  &v_0 \pi^\dagger        &-v_4 \pi^\dagger   \\
  0                     &-\frac{u}{2}       &-v_4\pi                &v_0\pi             \\
  v_0\pi                &-v_4\pi^\dagger    &\frac{u}{2}+\Delta'    &\gamma_1           \\
  -v_4\pi               &v_0\pi^\dagger       &\gamma_1               &-\frac{u}{2}+\Delta'
  \end{array}
  \right)&
  \begin{array}{cc}
  \gamma_0&=-2.61 \textrm{eV} \\ 
  \gamma_1&=.361 \textrm{eV} \\ 
  \gamma_4&=.138 \textrm{eV} \\ 
  \Delta'&=.015 \textrm{eV} \\
\end{array}
\label{s1}
\end{align}
\end{widetext}
where the basis in the $K$ valley consists of the wavefunction weight on the four lattice sites in the bilayer graphene unit cell $(\phi_A,\phi_{B'},\phi_B,\phi_{A'})$.
Here $\pi=p_x- i p_y$ and $\pi^{\dagger}=p_x+ip_y$ are momentum operators, and $u=\phi_2-\phi_1$ is the potential difference across the bilayer induced by the perpendicular electric field. Velocities are defined in terms of the monolayer graphene lattice constant, $a=2.46 \textrm{\AA}$, as $v_0=\frac{\sqrt{3}}{2}a\gamma_0/\hbar=8.44\times 10^5$m/s, $v_4=\frac{\sqrt{3}}{2}a\gamma_4/\hbar=4.47 \times 10^4$.  We use values of the tight binding parameters from recent \textit{ab inito} calculations\cite{Sjung_accurate_2014} shown in Eq. \ref{s1}. Results for valley $K'$ can be obtained by a 3D inversion, which exchanges $K \leftrightarrow K'$, $u \leftrightarrow -u$, $B \leftrightarrow A'$, and $B' \leftrightarrow A$.


To extend this Hamiltonian to the case of large perpendicular magnetic field, we introduce creation and annihilation operators for the scalar Landau level wavefunctions localized on each lattice site, defined as $\hat a \equiv \ell_B (q_x - iq_y)$ and $\hat a^{\dagger}\equiv\ell_B (q_x - iq_y)$ where $q_i\equiv k_i-\frac{e}{c}A_i$ and $\vec A$ is the magnetic vector potential.  The operators operate on scalar Landau level wavefunctions such that $\hat a|n\rangle=\sqrt{n}|n-1\rangle$ and $\hat a^\dagger|n\rangle=\sqrt{n+1}|n+1\rangle$.  The Hamiltonian in valley $K$, for example, then becomes
\begin{widetext}
\begin{equation}
  \hat H_K^B=  \hbar\omega_0\left(
  \begin{array}{cccc}
  \frac{u}{2\hbar\omega_0}           &0                      &\hat a^\dagger        &-\frac{\gamma_4}{\gamma_0} \hat a^\dagger   \\
  0                     &-\frac{u}{2\hbar\omega_0}           &-\frac{\gamma_4}{\gamma_0}\hat a                &a             \\
  \hat a      &-\frac{\gamma_4}{\gamma_0}\hat a^\dagger    &\frac{u}{2\hbar\omega_0}+\frac{\Delta'}{\hbar\omega_0}    &\frac{\gamma_1}{\hbar\omega_0}           \\
  -\frac{\gamma_4}{\gamma_0} \hat a               & \hat a^\dagger           &\frac{\gamma_1}{\hbar\omega_0}           &-\frac{u}{2\hbar\omega_0}+\frac{\Delta'}{\hbar\omega_0}
  \end{array}
  \right)
\label{hamB}
\end{equation}
\end{widetext}
where the monolayer graphene cyclotron energy is  $\hbar \omega_0=\frac{\hbar v_0\sqrt{2}}{\ell_B}\approx 30.6 \sqrt{\textrm{B}_{\perp}/\textrm{Tesla}} ~\textrm{meV}$.  As at $B=0$, the $H_{K'}$ follows from inversion, while spin enters only as an additional Zeeman energy $\sigma E_Z$.

The eigenstates of Eq.~\eqref{hamB} take the general form
\begin{align}
\ket{\xi N } = \sum_n (c_{\xi N; A}^n \ket{n}, c_{\xi N; B'}^n \ket{n}, c_{\xi N; B}^n \ket{n},  c_{\xi N; A'}^n \ket{n} ) , \label{eq:c_def}
\end{align}
where $\xi$ labels the valley (henceforth denoted $\xi = \pm$), $N$ labels the orbital quantum number, and the $\ket{n}$ are the oscillator states of $\hat{a}$, equivalent to the conventional quadratic-band LL-wavefunctions.  The coefficients are then determined by the band-structure.

The $N\geq2$ orbitals have energy $E_N\approx \hbar \omega_c\sqrt{N(N-1)}$, where
$\hbar \omega_c\approx \frac{3a^2\gamma_0}{2\ell_B^2\gamma_1}\gamma_0$, while the $N=0$ and $1$ orbital are nearly degenerate, both having zero energy for  $u=\gamma_4=\Delta'=0$.
$\gamma_4$, $\Delta'$, and finite $u$ all weakly lift this degeneracy, but still leave an eight fold near-degeneracy between states of different orbital, spin, and valley quantum numbers.  The ZLL is well separated from the $N=\pm2$ states at $E_{\pm2}\approx\pm\sqrt{2}\hbar\omega_c\approx\pm113 \textrm{meV}$. While $N\geq 2$ LLs have support on all four sublattices in the unit cell, the $N = 0, 1$ eigenstates vanish on one or more sublattices:
\begin{align}
  |- 0\rangle    &=\left(|0\rangle,0,0,0\right)  \label{eq:4bandPsi1} \\
  |- 1\rangle    &=\left(c_A|1\rangle,0,c_{B} |0\rangle, c_{A'} |0\rangle\right) .
  \label{eq:4bandPsi2}
\end{align}
Defining the layer polarization as $\alpha\equiv |c_A|^2-|c_{B'}|^2+|c_B|^2-|c_{A'}|^2$, we note that within the 4 band model the $N = 0$ orbital is fully layer polarized ($\alpha_0 = 1$) but the $N=1$ is  not;  using the tight binding parameters above we find $\alpha_1 \approx .63$ at $B_\perp=31$T. Wavefunctions in the opposite valley have correspondingly opposite layer polarization.

Throughout our experiment $u / \hbar \omega_c \ll 1$, and to leading order the single-particle energies of the ZLL are
\begin{align}
H^{(1)} = -E_Z \sigma + \Delta_{10} N -  \xi \frac{u}{2} \alpha_N.
\label{eq:Ezll}
\end{align}
Here, as in the main text, $E_Z$ is the Zeeman energy, $\sigma = \pm \tfrac{1}{2}$ denotes the spin projection along the direction of the applied field, $\Delta_{10} \approx \hbar \omega_c (2 \frac{\gamma_4}{\gamma_0} + \frac{\Delta'}{\gamma_1})$ is the single particle orbital splitting, $u=\phi_2-\phi_1$ is the $E$-field induced potential difference across the bilayer, $\alpha_N$ is the layer polarization of the orbital, and $\xi = \pm $ indexes the valley.
The single particle levels are shown in Fig. 3A of the main text. At $B=31$T,  $\Delta_{10}=9.7$ meV, $E_Z=3.58$ meV, and $\{\alpha_0, \alpha_1\} = \{1, 0.63\}$ as follows from Eq. \ref{hamB}.
The large splitting to the higher Landau levels ensures that they are not involved in any $u$- or $E_Z$-tuned phase transitions.

\subsection{Coulomb Hamiltonian}
The Coulomb interactions decompose into a dominant isospin SU(4)-symmetric part and subleading capacitive and valley anisotropies:
\begin{align}
\mathbf{H^{(2)}}&\mathbf{= H^{SU(4)} + H^{\mathrm{c_0}} + H^{V}}
\end{align}
We now discuss these terms in turn.

\subsubsection{$\mathbf{H^{SU(4)}}$}
\emph{Screened Coulomb interaction.}
The bare Coulomb interaction is screened by the surrounding hBN dielectric, the proximal metallic gates, and filled LLs below the ZLL of the BLG itself.
The screening due to the hBN dielectric is incorporated as a dielectric constant in the Coulomb scale
\begin{equation}
E_C = \frac{e^2}{4 \pi \epsilon^\parallel_{\textrm{BN}} \ell_B} \sim 8.58 \sqrt{\textrm{B}/\textrm{T}}\textrm{meV},
\label{eq:Ec}
\end{equation}
assuming $\epsilon^\parallel_{\textrm{BN}} \approx 6.6 \epsilon_0$\cite{Sgeick_normal_1966,Sohba_first-principles_2001}. The metallic gates, each at distance $D \approx 20$nm from the bilayer, exponentially screen the interaction when $r \gg D$.  The 2D Fourier transform of the gate-screened potential is
\begin{align}
V(k)= \frac{2 \pi}{k} \tanh(k D )
\label{eq:bareV}
\end{align}
in units of $E_C$ and $\ell_B$.  Here we neglect the finite width of the bilayer itself, which is an order of magnitude smaller than $\ell_B$; it will be reincorporated as a capacitive energy below.

Since we wish to work with a model projected into the ZLL, we must account for the residual response of the other LLs, colloquially referred to as ``LL-mixing''.
The dimensionless parameter controlling their  response is $\frac{E_C}{\hbar \omega_0} \sim 3.14 / \sqrt{B/\textrm{T}} \approx 0.56$ at $B=31$T, which is comparable to values in GaAs.
Motivated by the large number of isospin flavors (four), the standard approach for BLG is the random phase approximation (RPA) ,\cite{Snandkishore_dynamical_2010, Sgorbar_broken_2012} in which we replace the bare Coulomb potential $V(k)$ with the effective potential
\begin{align}
V_{\textrm{eff}}(\omega, k) = \frac{V(k)}{1 + V(k) \Pi(\omega, k)},
\end{align}
where $\Pi$ is the polarization response. The RPA result is then further approximated by the static $\omega = 0$ value.
However,  RPA calculations have only been reported for the PH-symmetric two-band model\cite{Sgorbar_broken_2012} at $\nu = 0$, which is not quantitatively correct at large electric fields or at the magnetic fields relevant for our experiment\cite{Ssnizhko_importance_2012}.
Moreover, Ref.~\onlinecite{Sgorbar_broken_2012} found that the static $\omega = 0$ approximation strongly overestimates screening (IQHE gaps were underestimated by a factor of three).  Thus even recalculating the RPA value with a four band model is unlikely to be quantitatively accurate, as it is not possible to incorporate $\omega$ dependence into ground state numerical methods like exact diagonalization or DMRG.
Thus, at present, there does not appear to be a satisfactory ``ab initio'' tool for quantitatively predicting the strength of screening.


For these reasons, we use a phenomenological model following the approach of Ref.~\onlinecite{Spapic_topological_2014}, taking
\begin{align}
V_{\textrm{eff}}(k) = \frac{V(k)}{1 + a V(k) \tanh(b k^2 \ell_B^2 ) 4 \log(4)  / 2 \pi }
\end{align}
This is motivated by the RPA form when approximating the polarization as $\Pi(k) = a \tanh(b k^2 \ell_B^2 )  4 \log(4) / 2 \pi $, with $V(k)$  given in Eq.\eqref{eq:bareV}.
If $a, b$ are chosen to match the low-$k$ and high-$k$ behavior of the $\nu = 0$ two-band RPA calculation,\cite{Sgorbar_broken_2012} one finds $ a_\textrm{RPA} \equiv \frac{E_C}{\hbar \omega_0}   $ and  $b_\textrm{RPA} \equiv  0.62$.
However, much more generally we must have $\Pi(k) \propto k^2$ at low-$k$ and $\Pi(k) \to \textrm{const}$ at high-$k$, as captured by the ansatz.
Quantum Hall calculations are only sensitive to the form of the interaction in the vicinity of $k \lsim \ell_B^{-1}$, so the magnitude of the low-$k$ behavior ($k^2$) forms a  one-parameter space of screening behaviors set here by the product $a b$.
Thus we fix $b = b_\textrm{RPA}$, and treat $a_{\textrm{scr}}$ as a phenomenological measure of the screening strength.





\emph{ZLL projected Hamiltonian}-- We then project the screened Coulomb interaction $V_{\textrm{eff}}(q)$  into the  eight  components of the ZLL.
For the moment we neglect the small, lattice-scale  valley anisotropies which break the valley-SU(2) symmetry;  these effects will be introduced as phenomenological couplings shortly.
The SU(4)-isospin symmetric interaction is
\begin{widetext}
\begin{align}
\mathbf{H^{SU(4)}} = \tfrac{1}{2} \int \frac{d^2q}{(2 \pi)^2}  \, n_{\textrm{ZLL}}(q) V_{\textrm{eff}}(q) n_{\textrm{ZLL}}(-q) + \Delta_{\textrm{Lamb}} \sum_{\sigma \xi } \hat{N}_{ \xi 1 \sigma}
\end{align}
\end{widetext}
Here  $n_{\textrm{ZLL}}(q) = \sum_{\xi\sigma } n_{\xi\sigma}(q)$ is the total density in the ZLL, which is a sum of the four isospin components, while $\hat{N}_{ \xi 1 \sigma}$ is the electron number in level $ \xi 1 \sigma$.
As explained in Ref.~\onlinecite{Sshizuya_structure_2012}, a shift $\Delta_{\textrm{Lamb}}$ between the $N=0, 1$ orbitals arises when projecting into the ZLL, since their Coulomb exchange with the filled LLs below the ZLL differs.
Under the approximation of particle-hole symmetry, it was shown that $\Delta_{\textrm{Lamb}} = \frac{1}{2} ( E^{(\textrm{ex})}_{00}  - E^{(\textrm{ex})}_{11}) < 0$, where $\frac{1}{2}  E^{(\textrm{ex})}_{NN}$ is the Coulomb exchange  per-electron when fully filling an $N$ level with a Slater-determinant. We evaluate this shift using  the screened interaction $V_\textrm{eff}$, and find a near perfect fit to $\Delta_{\textrm{Lamb}}\simeq -\frac{0.2 E_C}{1 + 2.73 a_{\textrm{scr}}}$.

The density $n_{\xi, \sigma}(q)$ contains a contribution from both the $N=0$ and $N=1$ levels.
The ratio between the Coulomb scale and the single-particle splitting between the orbitals is
\begin{equation}\frac{E_C}{\Delta_{10} + \Delta_{\textrm{Lamb}} + \xi \tfrac{u}{2} (\alpha_0 - \alpha_1))} \approx \frac{E_C}{\Delta_{10} + \Delta_{\textrm{Lamb}}}\approx12.\label{eq:kappa}\end{equation}
Thus, unlike the levels outside the ZLL, it is not well-justified to project the interaction into only one of the two $N$ orbitals; we must keep both.

\emph{BLG form factors}--
For completeness, we explain how the BLG ``form factors''   $\mathcal{F}_
{NM}$ can be used to compute the Coulomb matrix elements projected into the Landau-level basis, which are required for the actual computations.
The density $n_{\sigma \xi}$ is not diagonal in the orbital index $N$, but instead involves ``orbital-mixing'' contributions:
\begin{align}
n_{\sigma \xi}(q) =  \sum_{N,M = 0, 1} \bar{\rho}_{\sigma \xi N M}(q) \mathcal{F}_{NM}(q)
\end{align}
Here $\bar{\rho}$ is a guiding-center density operator, which in the Landau gauge reads
\begin{align}
\bar{\rho}_{\sigma \xi N M }(q) \equiv \sum_k e^{-i k q_x \ell_B^2 } \psi^\dagger_{\sigma \xi N}(k + q_y / 2) \psi_{\sigma \xi M }(k - q_y / 2).
\end{align}
The form factor $\mathcal{F}$ is expressed in terms of the conventional quadratic-band form factors $F_{nm}$ as
\begin{align}
\mathcal{F}_{NM}(q) = \sum_{A_i, m, n} \bar{c}_{\xi N; A_i}^n F_{nm}(q) c_{\xi M; A_i}^m.
\end{align}
where $A_i = A, B, A', B'$ are the sublattices and $c$ are the wavefunction amplitudes defined in Eq.\eqref{eq:c_def}.
The $\mathcal{F}_{NM}$ are independent of $\xi$, since to leading order in $u$ the $\xi = \pm$ wavefunctions differ only by a permutation of the sites.

In the four-band model, we refer to Eqs.\eqref{eq:4bandPsi1}-\eqref{eq:4bandPsi2} to find
\begin{align}
\mathcal{F}_{00} = F_{00}, \quad \mathcal{F}_{01} = c_A F_{01}, \quad \mathcal{F}_{11} =c_A^2 F_{11} + (c_{A'}^2 + c_{B}^2) F_{00}.
\end{align}
It is thus convenient to parameterize the interaction by $\cos^2 \Theta = c_A^2, \sin^2 \Theta = c_{A'}^2 + c_{B}^2$ , where $\Theta \approx 0.44$ at $B=$31T.
At low perpendicular magnetic fields, $\Theta \to 0$ and the problem reduces to the two-band model
\begin{align}
\mathcal{F}_{00} = F_{00}, \quad \mathcal{F}_{01} = F_{01}, \quad \mathcal{F}_{11} = F_{11},
\end{align}
equivalent to the two lowest Landau-levels of the conventional quadratic-band QHE.

\subsubsection{$\mathbf{H^{\mathrm{c_0}}}$ and $\mathbf{H^{V}}$}
Lattice scale effects at order $a/ \ell_B$ generate valley-SU(2) breaking perturbations sensitive to the details of the lattice and orbital structure, so must be treated phenomenologically\cite{Skharitonov_canted_2012, Ssodemann_broken_2014}.
The first anisotropy is a capacitive energy due to the finite thickness of the BLG ($d$), which modifies the interlayer Coulomb interaction from $V(q) \to V(q) e^{-q d}$. This perturbation is smaller than the long range Coulomb ($H^{SU(4)}$) by a factor of $d / \ell_B \ll 1$, but the $q = 0$ part is unscreened, so for simplicity we retain the Hartree-type capacitive charging energy $E^{c_0} = \frac{1}{8 c_{0} }(N_t - N_b)^2$, where $N_{t/b}$ is the charge on the top / bottom layer, evaluated using $\alpha_1$.
Writing the capacitance of the BLG as $c_0 = \frac{\epsilon_{\textrm{BLG} A}^{\perp} }{d}$, we can manipulate the expression to obtain
\begin{align}
\mathbf{H^{\mathrm{c_0}}} = N_\Phi E_C \frac{d}{\ell_B} \frac{\epsilon_{\textrm{BN}}^{\parallel}}{\epsilon_{\textrm{BLG}}^{\perp}}\frac{(\nu_t - \nu_b)^2}{4}.
\end{align}
The ratio of  dielectric constants arises because we included ${\epsilon_{\textrm{BN}}^{\parallel}}$ in $E_C$.
The dielectric constant $\epsilon_{\textrm{BLG}}^{\perp}$ should be the same one used for converting the applied gate voltage $p_0 / c$ to the inter-layer bias $u$.

Following Refs.~\onlinecite{Skharitonov_canted_2012,Ssodemann_broken_2014}, the remaining valley anisotropies are incorporated as short-range ($\delta$-function) interactions which preserve spin $SO(3)$ and valley $U(1) \times U(1) \rtimes \mathbb{Z}_2$.
In contrast to monolayer graphene, symmetry considerations allow a rather large space of possible perturbations since they can take an arbitrary form in the $N = 0, 1$ orbital index. Here we will assume the perturbations can be expressed using local operators that depend only on the isospin, e.g. $O^{\mu \nu}(r) = \psi^\dagger_{\xi, \sigma}(r) \tau^{\mu}_{\xi, \xi'} \sigma^\nu_{\sigma, \sigma'} \psi_{\xi', \sigma'}(r)$, where $\psi_{\xi, \sigma} = \sum_N \psi_{\xi, N, \sigma}(r)$ and $\tau, \sigma$ are Pauli operators.
This assumption is reasonable when $\Theta \to 0$, since the anisotropies arise from lattice structure and the $N=0, 1$ orbitals sit on the same site in this limit. While this approximation neglects the $\Theta\neq0$ corrections due to the finite polarization of the N=1 levels, we note that the capacitance term $\mathbf{H^{\mathrm{c_0}}}$ captures at least some of these.

We express the anistropy energy using the total valley number density $n_{\pm}(r)$ and the valley spin density $\mathbf{S}_{\pm}(r) = \frac{1}{2} \sum_{MN ab} \psi^\dagger_{ \pm M a}(r) \mathbf{\sigma}_{a b} \psi_{ \pm N b}(r)$.
In addition to the total density (which preserves SU(4)), the most general symmetry preserving perturbation is
\begin{widetext}
\begin{align}
\mathbf{H^{V}} &= E_C \frac{d}{\ell_B} 2 \pi \ell_B^2 \int  d^2r  \left[- g_z n_+(r) n_-(r) + g_{\perp} \mathbf{S}_+(r) \cdot \mathbf{S}_-(r) \right]
\label{eq:hv}
\end{align}
\end{widetext}
Note that previous treatments\cite{Skharitonov_canted_2012} have assumed interactions of the type $\frac{u_\perp}{2} (\tau^x \tau^x + \tau^y \tau^y)$ and $\frac{u_z}{2} \tau^z \tau^z $.
These forms are in fact equivalent: using the anti-symmetry of the fermions, we have $E_C \frac{d}{\ell_B} g_\perp = - 4 u_\perp, E_C \frac{d}{\ell_B} g_z =  u_\perp  + 2 u_z$.



\subsection{Evaluation of $H^{(2)}$}

\begin{table*}[t]
\begin{center}
\caption{Model parameters}
\label{tab:parameters}
\begin{tabular}{|c|c|c|c|}
\hline
\textbf{Parameter} & \textbf{Expression} & \textbf{Value} & \textbf{Source} \\ \hline
$\epsilon_{BN}^\parallel$       & --   & 6.6 $\epsilon_0$                                                                           & literature\cite{Sgeick_normal_1966,Sohba_first-principles_2001}   \\ \hline
$\epsilon_{BN}^\perp   $        & -- & 3.0 $\epsilon_0$                                                                             & measured (Sec. \ref{sec:mease})     \\ \hline
$\Delta_{10}$                   &  $E_{\xi1\sigma}(u=0)-E_{\xi0\sigma}(u=0)$   &  9.7 meV                                           & derived from band structure       \\ \hline
$\alpha_{1}$                    &  $\phi_A^2-\phi_{B'}^2+\phi_B^2-\phi_{A'}$ & 0.63                                                 & derived from band structure\cite{Sjung_accurate_2014}      \\ \hline
$\hbar\omega_c$                 &  $\frac{3a^2\gamma_0}{2\ell_B^2\gamma_1}\gamma_0$ & 80 meV                                        & derived from band structure\cite{Sjung_accurate_2014}      \\ \hline
$E_C$                           &  $\frac{e^2}{4\pi \epsilon_{BN}^\parallel\ell_B}$&  47.3 meV                                      & derived from $\epsilon_{BN}^\parallel$\\ \hline
$a_{RPA}$                       & $\frac{E_C}{\hbar \omega_c}$ &  .42                                                               & derived from $E_C$ and $\hbar \omega_c$\\ \hline
$\zeta$                         & $\frac{E_C }{2} ( - E_{00}^{(\textrm{ex})} +
E_{11}^{(\textrm{ex})} +  2 E_{01}^{(\textrm{ex})} )= -\frac{0.315 E_C}{1 + 2.52 a_\textrm{scr}}$ & -8.8  meV                       & calculated (Eq. \ref{zeta})    \\ \hline
$\Delta_{\textrm{Lamb}}$        &  $\frac{E_C}{2} ( E_{00}^{(\textrm{ex})}
- E_{11}^{(\textrm{ex})} ) = -\frac{0.2 E_C}{1 + 2.73 a_{\textrm{scr}}}$&  -5.4 meV                                                 & calculated (Eq. \ref{lamb})      \\ \hline
$g_\perp$                       & -- & 0.69                                                                                         & estimated from data (Sec. \ref{sec:estgperp}) \\ \hline
$g_z$                           & --  &  0.45                                                                                       & \textrm{fit from phase transitions (Sec. \ref{sec:fits})}      \\ \hline
$\epsilon_{BLG}^\perp  $        &  --   & 2.8                                                                                       & \textrm{fit from phase transitions (Sec. \ref{sec:fits})}      \\ \hline
$a_{scr}$                       & -- & .28$\approx$.67$a_{RPA}$                                                                                          & \textrm{fit from phase transitions (Sec. \ref{sec:fits})}\\ \hline
\end{tabular}
\end{center}
\end{table*}

Our model depends on a number of parameters, summarized in Tab.~\ref{tab:parameters}. Some are known from the literature (e.g., $\epsilon_{\textrm{BN}}^\parallel$ and the tight binding parameters shown in Eq. \ref{s1}), some can be derived directly from these assumed values (e.g., $\Delta_{10}, \alpha_1, E_C$), and some  follow from theory (e.g., $\zeta, \Delta_{\textrm{Lamb}}$). In addition, there are several phenomenological parameters which we constrain from experiment, namely $\epsilon_{\textrm{BN}}^\perp, g_\perp, g_z, a_{\textrm{scr}}, \epsilon_{\textrm{BLG}}^\perp$.

To get a handle on these parameters, we begin with a Hartree Fock approximation for $H^{(2)}$, which allows us to conveniently estimate the predicted location of the transitions within our model.

\subsubsection{Hartree Fock approximation for $H^{(2)}$}
	\begin{figure*}[t]
	\begin{center}
\includegraphics[width=.75\columnwidth]{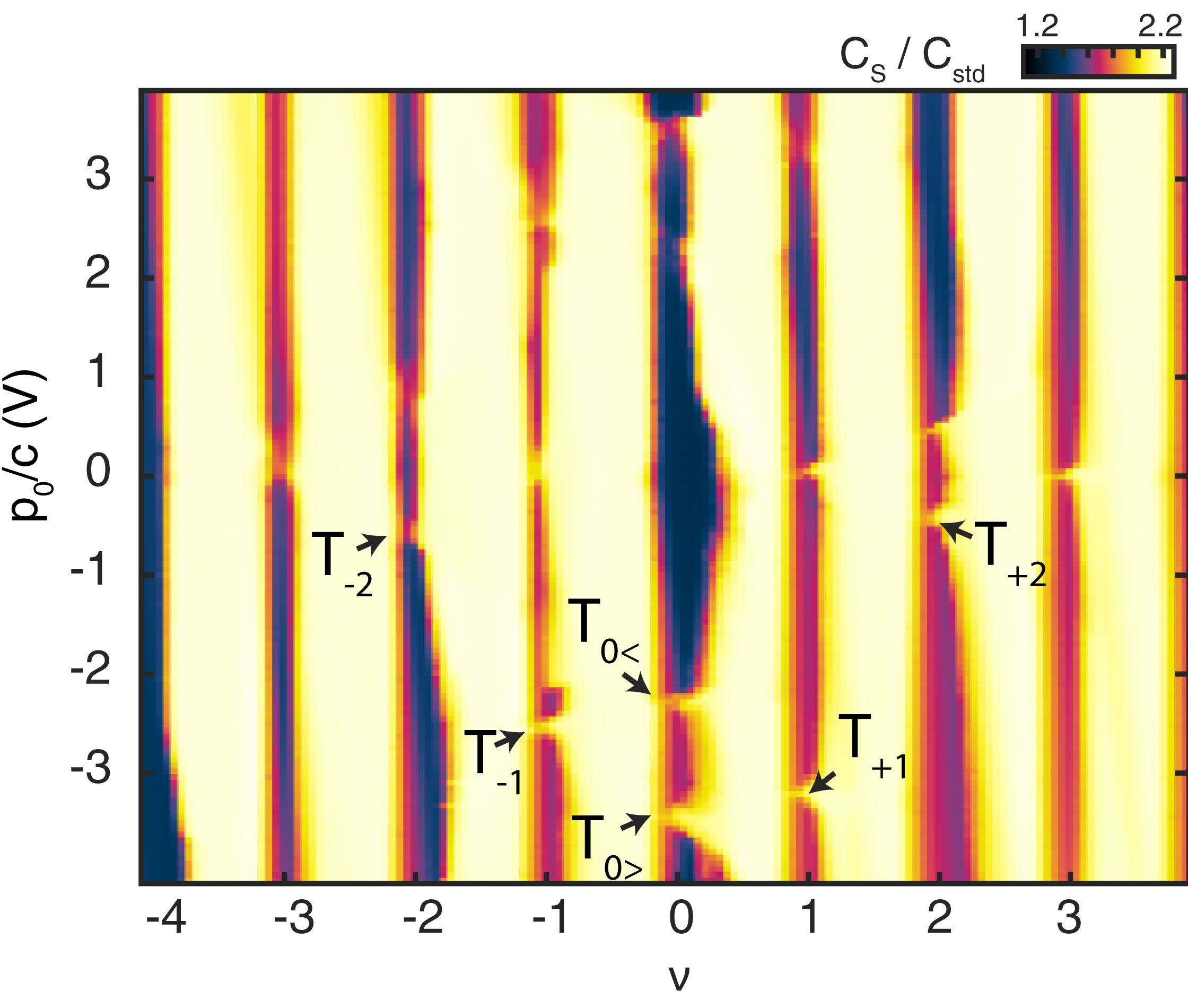}
		\caption{$C_S$ with phase transitions labeled for reference.}
		\label{fig:labels}
	\end{center}
 \end{figure*}

 Within the Hartree-Fock approach, we assume that the ground-state at integer filling is a Slater determinant which  successively  fills orbitals $\xi N \sigma$ according to the proposed phase.
We neglect the possibility of valley off-diagonal coherence (e.g., non-zero $\langle \psi^\dagger_{\xi}  \psi_{\xi'} \rangle \neq 0, \xi \neq \xi'$).
While such phases have been predicted to occur, for instance in a very narrow range of $p_0$ at $\nu = -3$,\cite{Slambert_quantum_2013} there is no evidence for them in the current data, since there appears to be a single direct transition at $p_0 = 0$.

The energy of $\mathbf{H^{SU(4)}}$ at integer filling can  then computed as an integral involving $\mathcal{F}_{NM}$ and $V_{\textrm{eff}}$ (see, for example, Ref. \onlinecite{Slambert_quantum_2013}). In contrast to earlier works, we use both the screened interactions and the form factors appropriate to the $B=31$T four-band model, as described above.
We interpolate the integer result to fractional $\nu$ by generalizing the first term in the interpolation of Fano and Ortolani\cite{Sfano_configuration-interaction_1986} to the multi-component case:
\begin{align}
\frac{E^{SU(4)}}{N_\Phi} \approx \frac{E_C}{2}  \sum_{\sigma \xi N, N'} \nu_{\sigma \xi N} E^{(\textrm{ex})}_{N, N'} \nu_{\sigma \xi N'} +  \Delta_{\textrm{Lamb}} \sum_{\sigma \xi} \nu_{\sigma \xi 1}
\label{eq:su4HF}
\end{align}
The entry $\frac{1}{2} E_{00}^{(\textrm{ex})}$ is the energy to fill a $N=0$ level; $\frac{1}{2} E_{11}^{(\textrm{ex})}$ the energy to fill a $N=1$ level; and $\frac{1}{2} E_{00}^{(\textrm{ex})} + \frac{1}{2} E_{11}^{(\textrm{ex})} +  E_{01}^{(\textrm{ex})}$ is the energy to fill both. The splitting between filling $N=0,1$ orbitals of the same isospin relative to two $N=0$ orbitals of opposite isospin is then $\zeta = \frac{E_C }{2} ( - E_{00}^{(\textrm{ex})} +  E_{11}^{(\textrm{ex})} +  2 E_{01}^{(\textrm{ex})} )$. Calculations were repeated for a range of $a_\textrm{scr}$, and we find
\begin{equation}
\zeta = -\frac{0.315 E_C}{1 + 2.52 a_\textrm{scr}}\label{zeta}\end{equation}
 gives an almost perfect interpolation of the result. The Lamb shift, meanwhile, is found to be
\begin{equation}
  \Delta_{\textrm{Lamb}}  = \frac{E_C}{2} ( E_{00}^{(\textrm{ex})} - E_{11}^{(\textrm{ex})} ) = -\frac{0.2 E_C}{1 + 2.73 a_{\textrm{scr}}}.\label{lamb}\end{equation}
 Note that this expression for the Lamb shift remains true beyond Hartree-Fock, so will also be used in our DMRG calculation.

The valley-anisotropies are off-diagonal in the valley index, so they reduce to a Hartree energy, e.g. $\langle n_+(r) n_-(r) \rangle \to \langle n_+(r) \rangle \langle n_-(r) \rangle$. Using $n_\pm(r) = \frac{1}{2 \pi \ell_B^2} \nu_{\pm}$, we arrive at the final expression

\begin{widetext}
\begin{equation}
\frac{E^{(2)}_{\textrm{H.F.}}}{E_C} = \frac{1}{2} \sum_{\xi N \sigma , N'} \nu_{\sigma \xi N} E^{(\textrm{ex})}_{N, N'} \nu_{\sigma \xi N'} + \Delta_{\textrm{Lamb}} \sum_{\sigma \xi } \nu_{\sigma \xi 1} + \frac{1}{4} \frac{d}{\ell_B} \frac{\epsilon^\parallel_{\textrm{BN}}}{\epsilon^\perp_{\textrm{BLG}}} (\nu_t - \nu_b)^2 - g_z \frac{d}{\ell_B} \nu_+ \nu_- + g_{\perp} \frac{d}{\ell_B} \mathbf{S}_+ \cdot \mathbf{S}_-
\label{eq:model}
\end{equation}
\end{widetext}

\subsubsection{Experimental determination of $\epsilon_{\textrm{BN}}^\perp$\label{sec:mease}} Since $\epsilon_{\textrm{BN}}^\perp$ determines the capacitance between the gates and BLG, we can measure $\epsilon_{\textrm{BN}}^\perp$ by fitting the Landau level spacing, as measured in the applied gate voltages $n_0/c$, to their known densities $n = \frac{\nu}{2 \pi \ell_B^2}$.
Data were taken at $p_0=0$ and $B_\perp=2T$.  Starting from the electrostatic model of Eqs. \ref{elec1}-\ref{elec2}, we ignore interlayer capacitance ($c_0\sim \infty$) to treat the bilayer as a single 2D electron system, and neglect the finite quantum capacitance, which is reasonable at low fields. Fitting the separation, in $n_0$, of two four fold degenerate LLs to $\Delta n_0=\epsilon_{\textrm{BN}}^\perp/d_{BN}\left(\Delta v_t+\Delta v_b\right)=\frac{4}{2\pi\ell_B^2}$, we find $\epsilon_{\textrm{BN}}^\perp \sim 3.0\pm.15$.

\subsubsection{Experimental estimate of $g_\perp$\label{sec:estgperp}}

	\begin{figure*}[ht!]
	\begin{center}
\includegraphics[width=4 in]{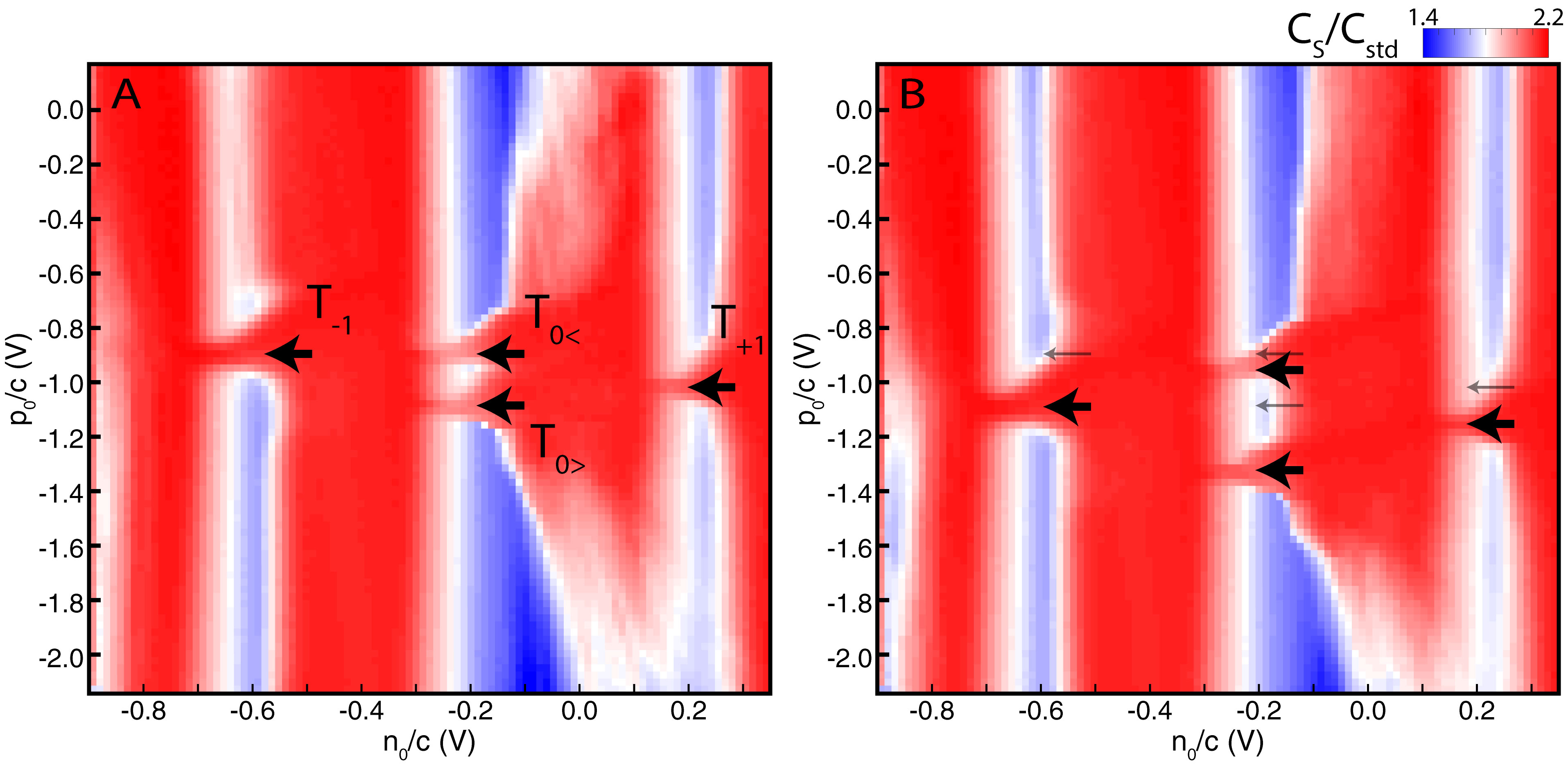}
		\caption{Tilted magnetic field dependence of integer transitions.
(\textbf{a}) Finite $p_0$ transitions at $\nu=0$ and $\nu=\pm1$ at $B_{tot}=B_\perp=15T$.
(\textbf{b}) Similar data at $B_{tot}=31T$ and $B_\perp=15T$.  Arrows indicate tilted and non-tilted positions of the phase transitions. The shifts are listed in Tab. \ref{tab:tilt}
}
		\label{fig:tilt}
	\end{center}
 \end{figure*}

By measuring the critical total $B$-field of the ferromagnet to canted antiferromagnetic transition at $\nu = p_0 = 0$, previous experiments\cite{Spezzini_critical_2014} have estimated that $B^{\textrm{tot}}_\ast = (2.42 \pm 0.21) B^\perp$. This implies $u_\perp = 0.14$meV/T, or in our parameterization, $g_\perp=\frac{u_\perp}{E_C}\frac{\ell_B}{d}=1.2$.
While this experiment shows the expected $B^\perp$-linear scaling for $B^\perp = 4-7$ T, we must be careful when extrapolating to higher magnetic fields.

To address this question, we analyze the tilted field measurements of four phase transitions at $B_\perp=15$T, as shown in Fig. \ref{fig:tilt}.  From the experimental data, we extract the shift in $\Delta p_0/c$ over the probed magnetic field range.  Each transition is expected to shift by $\Delta u=\frac{\overline{\sigma} \Delta E_Z}{\overline{\alpha}}$, where $\overline\alpha$ is the effective change in layer polarization across the transition and $\overline\sigma$ is the effective change in spin, and $\Delta E_Z = 0.116 \Delta B^{\text{tot}}$ meV/T the change in Zeeman energy.
 Solving for $\overline\sigma$, we find
\begin{equation}
\overline\sigma=\frac{c}{2c_0}\Delta(p_0/c){\overline{\alpha}}= .0086\frac{\epsilon_{\textrm{BN}}^\perp}{\epsilon_{\textrm{BLG}}^\perp}\frac{\Delta(p_0/c)}{\Delta E_Z}{\overline{\alpha}},
\label{eq:tilt}
\end{equation}
where $0.0086 = \frac{0.335\textrm{nm}}{39\textrm{nm}}$ is the ratio of geometric capacitances. When calculating $\bar{\alpha}$, we must remember that $\alpha_1=0.8$  at $B_\perp$=15T, as extracted from band structure parameters\cite{Sjung_accurate_2014}.
In the absence of antiferromagnetism, all transitions considered involve reversal of one full electron spin, so that $\overline\sigma=1$.  Canting of the spins due to $g_\perp$ can reduce the effective spin; however, $\overline \sigma>1$ is unphysical as it would imply a larger than unity spin per electron. Taking into account experimental error, each tilted field data point thus imposes a lower limit on $\epsilon_{\textrm{BLG}}^\perp$, the most stringent of which is $\epsilon_{\textrm{BLG}}^\perp > 2.57$.  As we will see below, this limit is ultimately consistent and does not influence the final fitted value of $\epsilon_{\textrm{BLG}}^\perp=2.76$.

Using this value for $\epsilon_{\textrm{BLG}}^\perp$, we can then calculate $\overline\sigma$ for each transition, shown in Table \ref{tab:tilt}.  Two of the four spin transitions are consistent with absence of canting ($\overline\sigma=1$), and two are consistent with strong canting ($\overline \sigma<1$). Additional data (not shown) shows that the phase transitions at $\nu = -2$ do not depend on in-plane magnetic field, suggesting that antiferromagnetism has not yet set in at this filling.

\vspace{7pt}
\begin{table}[]
\centering
\begin{tabular}{|c|c|c|c|c|}
\hline
Transition   &$\Delta(p_0/c)$, \textrm{V}    &$\overline{\alpha}$    &$\epsilon_{blg}^{min}$&$\overline\sigma$, $\frac{\epsilon_{blg}}{\epsilon_{BN}^\perp}=\frac{2.76}{3}$\\\hline
$T_{-1}$      &.21(.025)                       & 1.0                   & 2.58 & 1.06 (.13) \\
$T_{0<}$       &.23(.025)                       & 0.9                   & 0.31  & .23 (.11)  \\
$T_{0>}$       &.05(.025)                       & 0.9                   & 2.57  & 1.05 (.11)  \\
$T_{+1}$      &.13(.025)                       & 0.8                   & 1.17  & .53 (.1)  \\\hline
\end{tabular}
\caption{Effective spin involved in phase transitions near $\nu=0$, extracted from the data shown in Fig. \ref{fig:tilt} and Eq. \ref{eq:tilt}}
\label{tab:tilt}
\end{table}

In a  phase characterized by net spin $S_+$ and $S_-$ in each valley, the threshold for canting in our model arises from the interplay between the Zeeman effect (which favors polarization) and $g_\perp$ (which favors canting) via
\begin{equation}
  E_Z^c=g_\perp E_c \frac{d}{\ell_B}(S_+ + S_-),
\end{equation}
where observation of $\overline\sigma<1$ at a phase transition implies that $E_Z < E_Z^c$ for one of the adjacent phases.
\begin{table}[]
\centering
\begin{tabular}{|c|c|c|c|}
\hline
$\nu$&Transition   &Canting? & Constraint \\\hline
-2  &   $T_{-2}$   &   no          &   $E_Z>g_\perp E_c \frac{d}{\ell_B}$\\
-1  &   $T_{-1}$    &   no          &   $E_Z>\frac{3}{2}g_\perp E_c \frac{d}{\ell_B}$\\
0   &   $T_{0<}$    &   yes         &   $E_Z<2g_\perp E_c \frac{d}{\ell_B}$\\
0   &   $T_{0>}$    &   no          &   $E_Z>g_\perp E_c \frac{d}{\ell_B}$\\
+1  &   $T_{+1}$    &   yes         &   $E_Z<\frac{3}{2}g_\perp E_c \frac{d}{\ell_B}$\\\hline
\end{tabular}
\caption{Constraints on $g_\perp$ implicit in the presence of absence of spin canting at five phase transitions. Transitions refer to Fig. \ref{fig:tilt}.  No tilted field dependence is ever observed at $\nu=-2$, consistent with full spin polarization.}
\label{tab:gperp}
\end{table}
The canting at $T_{0<}$ and the absence of canting at both $T_{-2}$ and $T_{0>}$ together constrain $0.52<g_\perp<1.04$.
$T_{-1}$ and $T_{+1}$, meanwhile, seem to give  contradictory constraints, $0.69<g_\perp$ and $0.69>g_\perp$.
The lack of consistency likely stems from the dependence of the $g_\perp$ anisotropy on the orbital filling: $T_{-1}$ and $T_{+1}$ involve different partial fillings of an isospin flavor, and $g_\perp$ need not be identical for different orbital combinations in the different valleys.  At $T_{-1}$, occupation is transferred from a $|0+\uparrow\rangle$ to a $|0-\downarrow\rangle$ state, while at $T_{-1}$ occupation is transferred from a $|1+\sigma\rangle$ to a $|1-\sigma'\rangle$ state.  As we discussed, this effect is not captured in Eq. \ref{eq:hv}, which treats orbital components equally.
While the precise value and possible orbital substructure of $g_\perp$ is critical for determining the precise dependence of the antiferromagnetism on filling, within our model we find $g_\perp$ has negligible impact on the locations of the $u$-tuned phase transitions, which are the focus of this work.  Thus we set $g_\perp = 0.69$, and defer a more full analysis of the $\nu$ dependence of the antiferromagnetism to future high resolution studies of the tilted field dependence of the phase transitions observed in $C_A$.

\subsubsection{Estimating $g_z, a_{\textrm{scr}}, \epsilon_{\textrm{BLG}}^\perp$ from the experimental data \label{sec:fits}}

In order to estimate $g_z, a_{\textrm{scr}}, \epsilon_{\textrm{BLG}}^\perp$, we compare the Hartree-Fock model with experiment.
Given the Hartree-Fock energy $E^{(\textrm{HF})}_{\nu_{\xi N \sigma}} = \langle H^{(1)} + H^{(2)} \rangle_{\nu_{\xi N \sigma}}$ of a filling sequence, the phase boundaries $u_\ast(\nu)$ are determined by equating $E^{(\textrm{HF})}_{\nu_{\xi N \sigma}} = E^{(\textrm{HF})}_{\nu'_{\xi N \sigma}}$. The layer bias $u_\ast$ is then related to the experimentally measured gate bias via $\frac{p_0}{c} = u  \frac{2 c_0}{c} = \frac{\epsilon^{\perp}_{\textrm{BLG}}}{\epsilon^\perp_{\textrm{BN}}} \frac{40\textrm{nm}}{0.335\textrm{nm}}$.	
Experimental data points are provided by the positions, in $p_0/c$, of the transitions $T_{-2}$, $T_{-1}$, $T_{0>}$, and $T_{+1}$ (see Fig. \ref{fig:labels} for the labeling scheme). In the absence of canting (we have assumed $g_\perp = 0.69$), the predicted phase boundaries are	
\begin{widetext}
\begin{align}
u_\ast(T_{-2})  &= \frac{2}{ 1 + \alpha_1}  \left( \zeta + \Delta_{\textrm{Lamb}} + \Delta_{10} + \frac{E_C}{4 \epsilon_{\textrm{BLG}}^{\perp} } \frac{d}{\ell_B} (1 + \alpha_1)^2  - E_C \frac{d}{\ell_B}  \left[   - g_z  +  g_{\perp} \frac{1}{4}  \right] \right)\label{eq:modelu1}\\
u_\ast(T_{-1})  &=   \left( \frac{E_C}{ \epsilon_{\textrm{BLG}}^{\perp} } \frac{d}{\ell_B} \left[ 1 + \alpha_1 \right] - E_C \frac{d}{\ell_B}  \left[   - 2 g_z  +  \frac{1}{2}  g_{\perp}  \right]   + E_Z   \right)\label{eq:modelu2}\\
u_\ast(T_{0>})  &=   \frac{2}{ 1 + \alpha_1}   \left( \zeta + \Delta_{\textrm{Lamb}} + \Delta_{10}  + \frac{E_C}{4 \epsilon_{\textrm{BLG}}^{\perp} } \frac{d}{\ell_B} 3 (1 + \alpha_1)^2  - E_C \frac{d}{\ell_B}  \left[   - 3 g_z  +  g_{\perp} \frac{1}{4}  \right]  + E_Z  \right)\label{eq:modelu3}\\
u_\ast(T_{+1})  &=   \frac{1}{ \alpha_1}   \left( \frac{E_C}{4 \epsilon_{\textrm{BLG}}^{\perp} } \frac{d}{\ell_B} \left[ (1 + 2 \alpha_1)^2  - 1 \right] + E_C \frac{d}{\ell_B}   \left[   2 g_z  - \frac{1}{2}   g_{\perp}  \right] +  E_Z   \right)
\label{eq:modelu4}
\end{align}
\end{widetext}
To compare to experimental values of $p_0/c$ for each transition, we convert $V(T_{i}) = u_\ast(T_i)\frac{2 c_0}{c}$. An fifth constraint arises from the tilt-field dependence of Eq.~\eqref{eq:modelu2}, which implies
\begin{align}
\frac{\partial V(T_{-1})}{\partial B^\textrm{tot}} =  \frac{2 c_0}{c}  g\mu_B\approx.00462\epsilon_{\textrm{BLG}}^{\perp}  \frac{\textrm{meV}}{\textrm{T}}
\end{align}

\begin{table}[]
\centering
\begin{tabular}{|c|c|c|c|}
\hline
Name&meas. value (error)&best fit& unit\\\hline
$T_{-2}$                                        &  0.68 (.04)   &0.63   & V\\
$T_{-1}$                                        &  2.59 (.05)   &2.57   & V\\
$T_{0>}$                                        &  3.48 (.05)   &3.6   & V\\
$T_{+1}$                                        &  3.27 (.05)   &3.2   & V\\
$\partial V(T_{-1})/\partial B^\textrm{tot}$    &  13 (2.5)     &12.7   & mV/T\\ \hline
\end{tabular}
\caption{Experimentally measured constraints.}
\label{tab:exppara}
\end{table}

Given  five experimental constraints in three unknowns, we do a least-squares fit for our overconstrained model.  The model is validated, simultaneously fitting all data points across a wide range of $\nu$, and providing physically reasonable values for $\epsilon_{\textrm{BLG}}^{\perp}=2.76$, $a_{\textrm{scr}}=0.29=0.5 a_{\textrm{RPA}}$, and $g_z=.082$.  The resulting phase diagram is plotted in Fig. 3c of the main text, lower panel, and similarly shows good quantitative agreement with most of the features of the experimentally measured phase diagram.

\subsection{Phase boundaries from iDMRG}
In order to reproduce the kink in $u_\ast(\nu)$ observed in experiment around $-3 < \nu < -1$, we replace the
Fano-Ortolani interpolation of the HF result (Eq.~\eqref{eq:su4HF}) with the infinite-DMRG calculation, which takes full account of correlations.
As before, we must compute the energies of the competing phases.
Since the tilt-field dependence shows that the spins are polarized across the transition, we label the isospin only by its valley $\xi = \pm$.
At the transition, density is transferred between isospin components. The cartoon picture is that in phase ``$001$'' (low $p_0$), the orbitals fill in order $+0, -0, +1$, while in phase ``$010$'' (higher $p_0$), the fill in order $+0, +1, -0$.
More generally, valley U(1)$\times$U(1) symmetry allows us to assign separately conserved fillings $\nu_{+}$, $\nu_-$ to the two valleys.
The two competing phases are defined by their fillings $\nu_{+}, \nu_-$ (we write $\tilde{\nu} = \nu + 2$ in this regime):
\begin{align}
(\nu_+, \nu_-)_{010} &= \begin{cases} (2 + \tilde{\nu}, 0) & \tilde{\nu} < 0 \\  (2, \tilde{\nu}) & \tilde{\nu} > 0 \end{cases} \\
(\nu_+, \nu_-)_{001} &= \begin{cases} (1, 1+\tilde{\nu}) & \tilde{\nu} < 0 \\  (1 + \tilde{\nu}, 1) & \tilde{\nu} > 0 \end{cases}
\label{eq:fillings}
\end{align}
Note that for fractional $\nu$ very close to the transition, there are presumably a multitude of phases in which  \emph{fractional} filling has transferred between the valleys, which we do not consider above. However, the current experiment isn't sensitive to these more delicate states, which are characterized by a lower energy scale, so they will  appear as a small rounding of the phase boundary $u_\ast$.

The iDMRG conserves U(1)$\times$U(1), so can be used to find the lowest energy state at filling $(\nu_+, \nu_-)$ under the Hamiltonian $H = H^{(1)} + H^{(2)}$.
While in principle the full Hamiltonian could be simulated in DMRG, for technical reasons it greatly simplifies matters to decompose the Hamiltonian as
\begin{widetext}
\begin{align}
H = \left[  \sum_\xi \hat{N}_{\xi, 1} \Delta_{10} +  \mathbf{H^{SU(4)}} \right]_{\textrm{DMRG}}  + \left[  \sum_{\xi, N} \frac{u}{2} \hat{N}_{\xi, N}  \xi \alpha_N \right]_{\textrm{P.T.}} + \left[ \mathbf{H^{\mathrm{c_0}}} + \mathbf{H^{V}} \right]_{\textrm{H.F.}}
\end{align}
\end{widetext}
The dominant part, $ \left[ \cdot \right]_{\textrm{DMRG}}$, will be evaluated in DMRG. The $u$-dependence $ \left[ \cdot \right]_{\textrm{P.T.}}$ is evaluated in first order perturbation theory, by evaluating $\langle \hat{N}_{\xi, N} \rangle$ using the ground states found in DMRG.
When $\alpha_1 =1$ first order perturbation theory is exact, since the layer polarization commutes with the Hamiltonian, and our tests indicate that more generally this introduces negligible error.
We make this approximation so that $u_\ast(\nu)$ can be determined from a single DMRG run at $u = 0$, rather having to re-run DMRG for each value of $u$.
Finally, $\left[ \cdot  \right]_{\textrm{H.F.}} $ is evaluated by neglecting inter-valley correlations, e.g. by taking $\langle n_+(r) n_-(r) \rangle \to \langle n_+(r) \rangle \langle n_-(r) \rangle $:
\begin{align}
\left[ \cdot  \right]_{\textrm{H.F.}}  = N_\Phi E_C  \frac{d}{\ell_B} \left[ \frac{1}{4} \frac{\epsilon^\parallel_{\textrm{BN}}}{\epsilon^\perp_{\textrm{BLG}}} (\nu_t - \nu_b)^2  +  \nu_+ \nu_-( - g_z + g_{\perp} \frac{1}{4}) \right]
\end{align}
Since one of the two isospins is always at integer filling, and hence is largely inert, the inter-valley correlations are expected to be small (on top of the already small scale $\frac{d}{\ell_B}$). For example, at $\nu_+ = \nu_- = 1$,  DMRG  shows the pair correlation is around $\langle n_+(r) n_-(r) \rangle \sim 0.8 \langle n_+(r) \rangle \langle n_-(r) \rangle $, a slight suppression from the Hartree value.
Thus while largely negligible for determining the phase boundary $u_\ast$, effects of this form could be  relevant for understanding the smaller energy scale governing spin physics, an interesting subject for future work.

	All parameters \emph{except} $a_{\text{scr}}$ are  those used or determined by the Hartree-Fock analysis. As will be discussed, we find that $a_{\text{scr}} = 0.22$ must be adjusted slightly to preserve the location of the $\nu = -2$ transition.

\paragraph{iDMRG numerics.}
In the infinite-DMRG method we place the quantum Hall problem $\left[ \cdot \right]_{\textrm{DMRG}}$ on an infinitely long cylinder of circumference $L$, for which we compute the ground state energy density including the full effect of correlations.
When both $2 > \nu_{+}, \nu_{-} > 0$ (for instance, in phase 001), it is necessary to keep \emph{four} ZLL components in the iDMRG, $+0, +1, -0, -1$. This is because even when $\nu_\xi = 1$,  isospin $\xi$ acts as a polarizable medium due to the very small splitting between the $N=0, 1$ orbitals.
While computationally very expensive,\cite{Szaletel_infinite_2015} the  ZLL orbital mixing is thus fully accounted for.

\begin{figure}[t]
	\begin{center}
\includegraphics[width=\columnwidth]{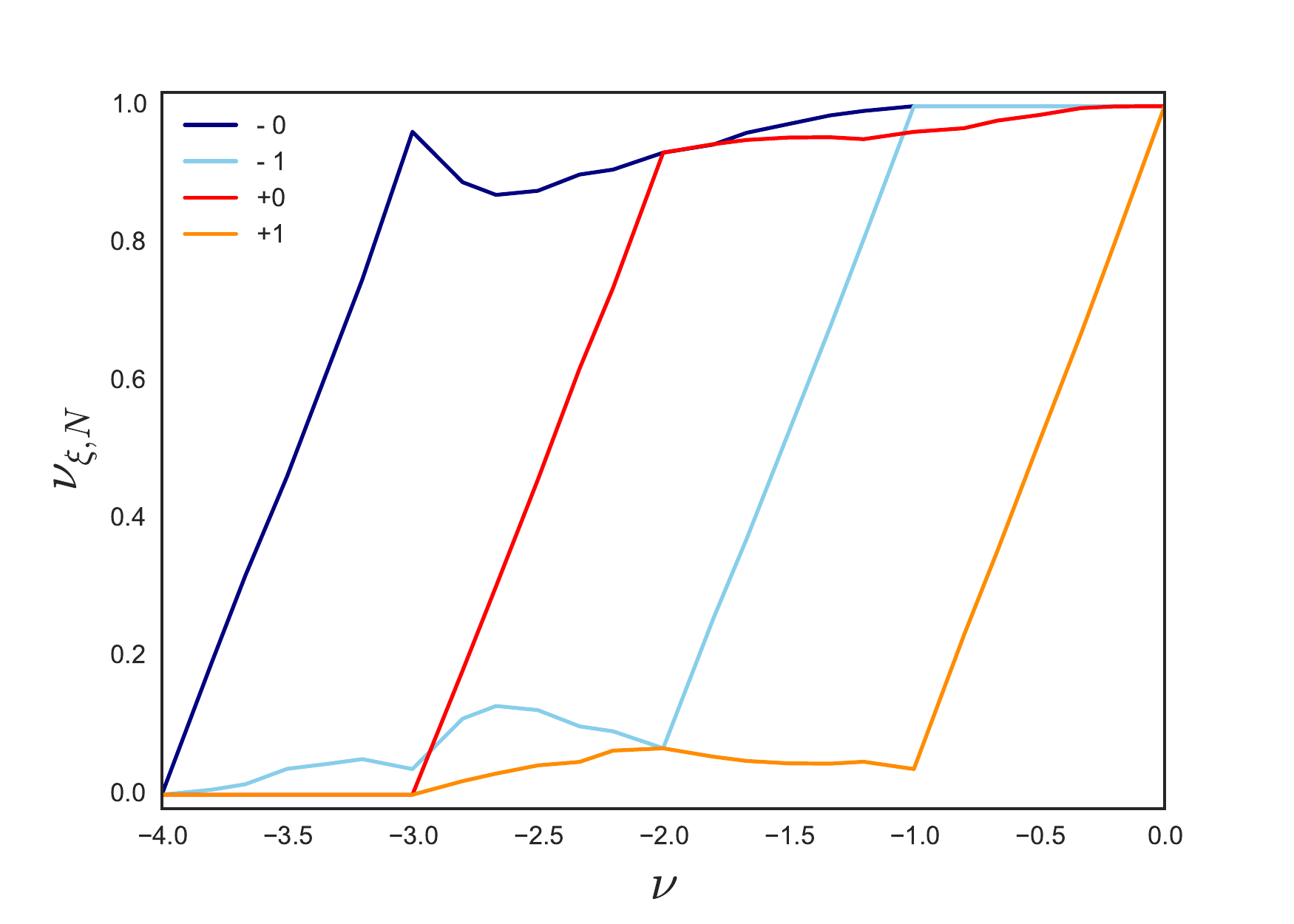}
		\caption{The densities $\nu_{\pm, N} = 2 \pi \ell_B^2 \langle n_{\pm, N}\rangle$  (we assume spin-polarization for simplicity) as the total filling $\nu$ is increased. $p_0 < 0$ is small, so the orbitals fill in the order $-0, +0, -1, +1$.
We see that while charge does fluctuate between $N=0, 1$ orbitals of the same isospin, this deviation is at most about 10\% of the filling.
This explains why blue/red ($N=0$) and cyan/orange ($N=1$) appear as a sharp contrast in our $C_A$ data.}
		\label{fig:fillings}
	\end{center}
 \end{figure}

In principle $H$ has a delicate $\nu$-dependence sensitive to all the fractional competing phases (which may be distinguished at the level of $10^{-3}$ or $10^{-4} E_C$), which would require finite-scaling analysis to fully resolve.
However, given the resolution of the present experiment, we focus on the much larger and slowly varying background (at the level of $10^{-1} E_C$).
For this purpose, we  work on cylinders of circumference $L = 16 \ell_B$ and use a DMRG-bond dimension of $\chi = 1600$, which results in an error in the energy per particle of around $10^{-4} E_C$, much smaller than the experimental features to be modeled.
Specifying the valley fillings $\nu_+, \nu_-$  according to Eq.~\eqref{eq:fillings}, iDMRG was used to compute the energy under $\left[ \cdot \right]_{\textrm{DMRG}}$ for the two competing phases at $1 + \tilde{\nu} \in \{0, 1/5, 1/3, 2/5, 1/2, 3/5, 2/3, 4/5, 1, \cdots, 2 \}$.

\begin{figure}[t]
\includegraphics[width=\columnwidth]{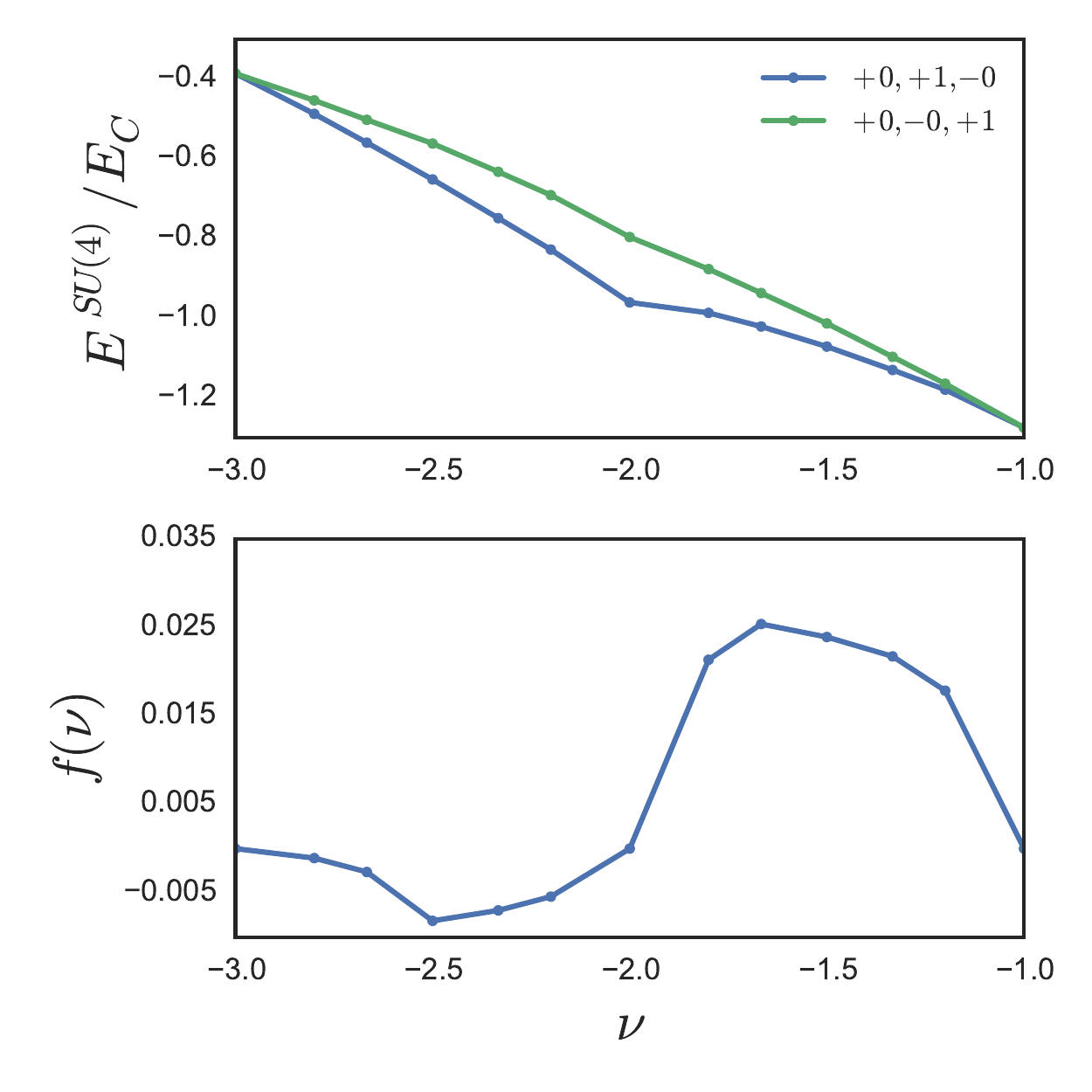}
\caption{
\textbf{a}) The SU(4)-symmetric part of the interaction energy for the two competing phases.
\textbf{b}) The difference between these interaction energies, after subtracting out the linear contribution.
\label{fig:EXamples}}
\end{figure}

Before discussing the resulting DMRG energies, we first address the issue of why blue/red and cyan/orange appear as distinct scales in the experimental $C_A$ data, which requires  that the $N=0, 1$ levels to largely fill sequentially rather than as a mixture.
Consider, for instance, the low $p_0 <0$ phase in which (naively) the orbitals fill in the order $-0, +0, -1, +1$. With interactions, charge can fluctuate between the $N=0, 1$ orbitals of the same valley.
In Fig.~\ref{fig:fillings}, we show the iDMRG result for the densities $\nu_{\pm, N} = 2 \pi \ell_B^2 \langle n_{\pm, N}\rangle$ as the total filling $-4 < \nu < 0$ is varied.
The deviation from the non-interacting expectation is at most around 10\%, explaining the sharp contrast.

In Fig.~\ref{fig:EXamples}a, we show the interacting part of the energy $E^{SU(4)}$  for both phases.
The difference $\zeta = E^{SU(4)}_{010}(\nu = -2) - E^{SU(4)}_{001}(\nu = -2)$ is slightly smaller than its Hartree-Fock value, so to preserve the location of the transition we adjust the screening to $a_{\textrm{scr}} = 0.22$ (this adjusts $\Delta_{\textrm{Lamb}}$ accordingly).
In Fig.~\ref{fig:EXamples}b, we show their difference after extracting out the linear part, $f(\nu) = E^{SU(4)}_{010}(\nu) - E^{SU(4)}_{001}(\nu) - \zeta (1 - |\nu + 2|) $.
The significantly different curvature on the two sides of $\nu = -2$ leads to the kink in the phase boundary discussed in the main text.

After calculating the   DMRG and  H.F. part of the energy at $u = 0$, the remaining polarization energy is determined from the DMRG expectation values $\frac{u}{2} \xi \alpha_N \langle n_{\xi N}\rangle$ shown in Fig.~\ref{fig:fillings}.
From this we compute the boundary $u_\ast(\nu)$ shown in the main text.

\newpage
\section*{Supplementary Data Figures}
\begin{figure*}[ht!]
	\begin{center}
\includegraphics[width=4.5in]{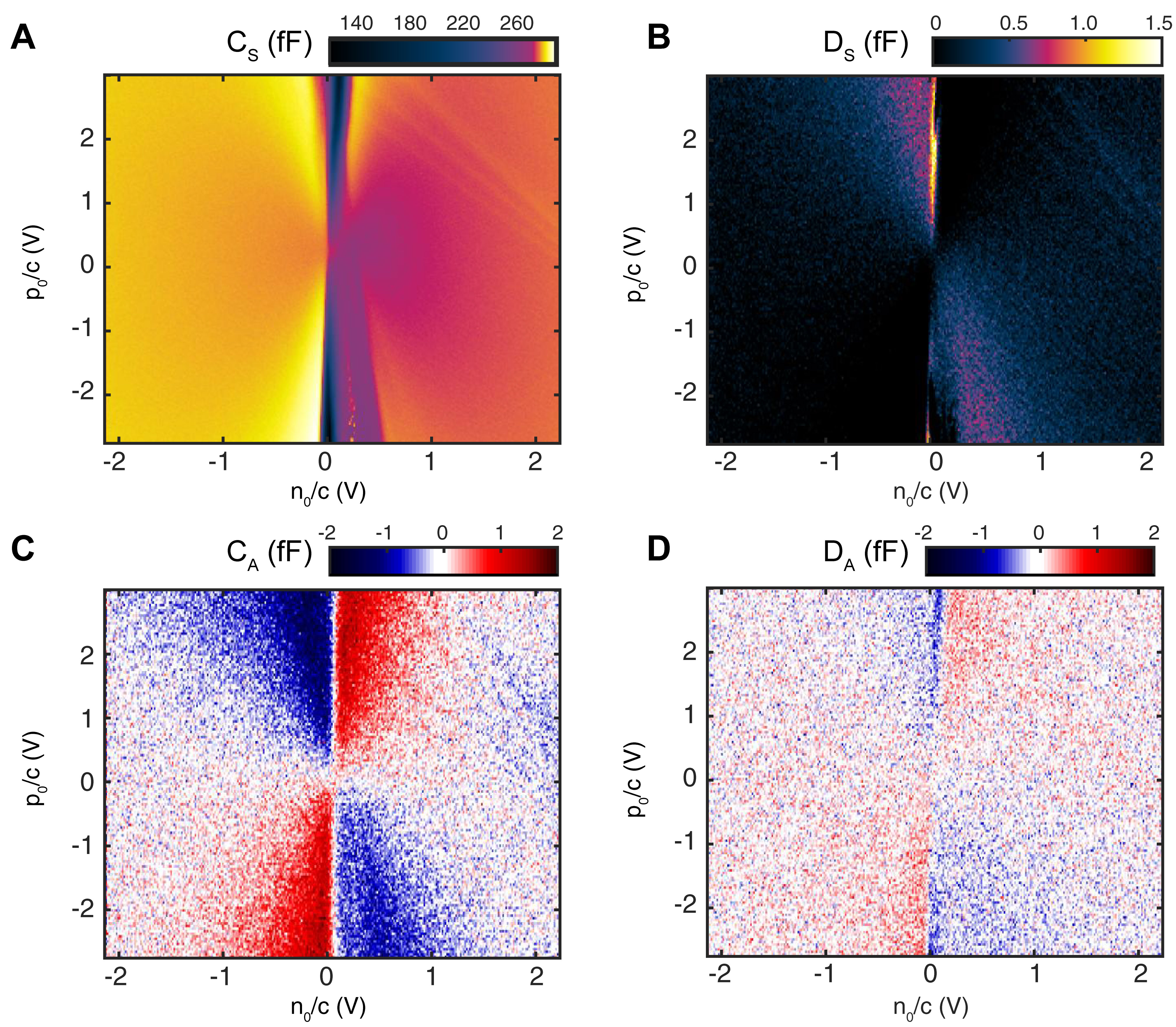}
		\caption{Capacitance and dissipation for the data set in Fig. 1 of the main text. Note that the scale in B is 1/100 of that in A.}
		\label{CapAndDis1}
	\end{center}
 \end{figure*}

\begin{figure*}[bp!]
	\begin{center}
\includegraphics[width=4.5 in]{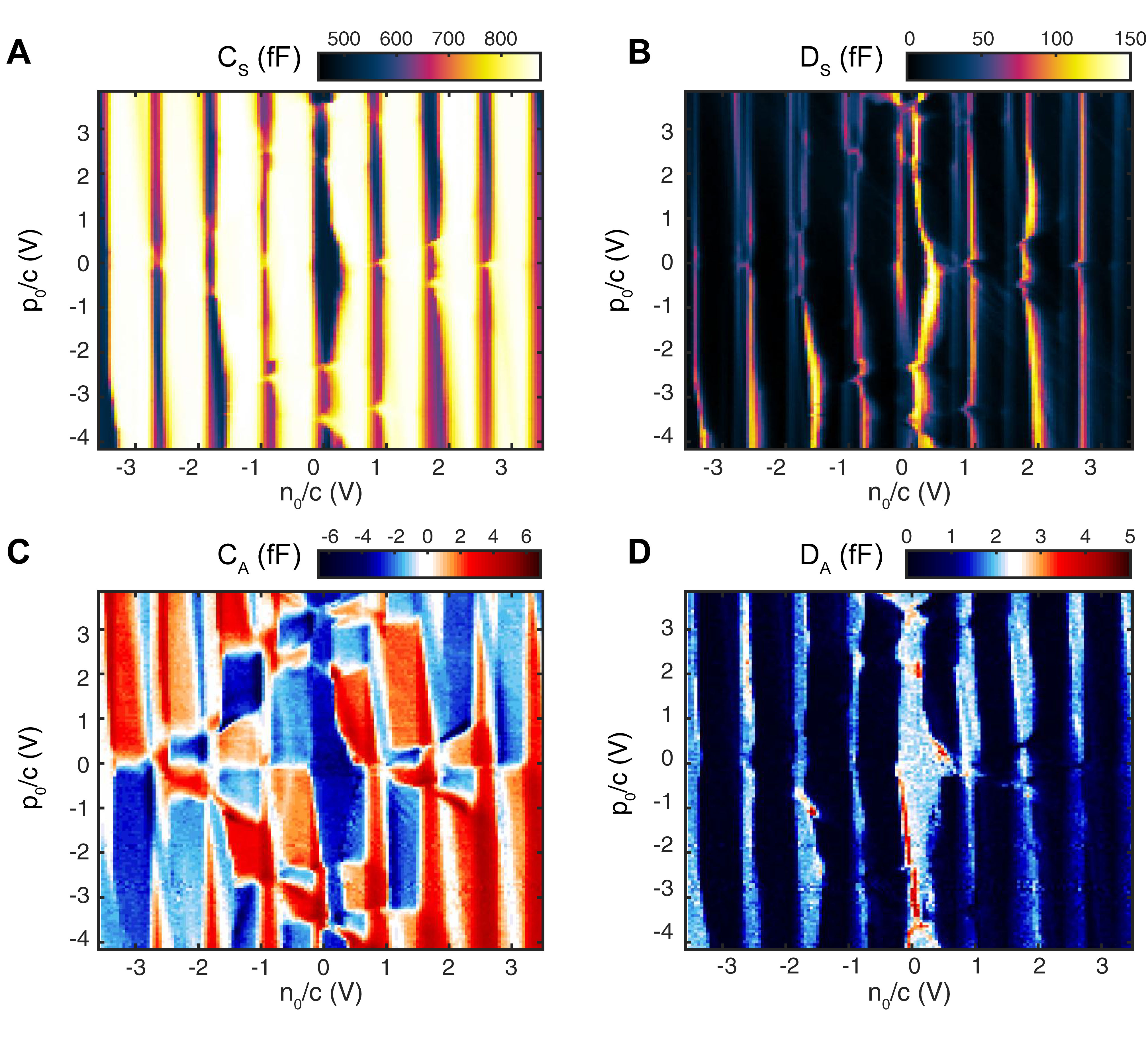}
		\caption{Capacitance and dissipation for the data set in Fig. 2 of the main text.}
		\label{CapAndDis2}
	\end{center}
 \end{figure*}

\begin{figure*}[ht!]
	\begin{center}
\includegraphics[width=4 in]{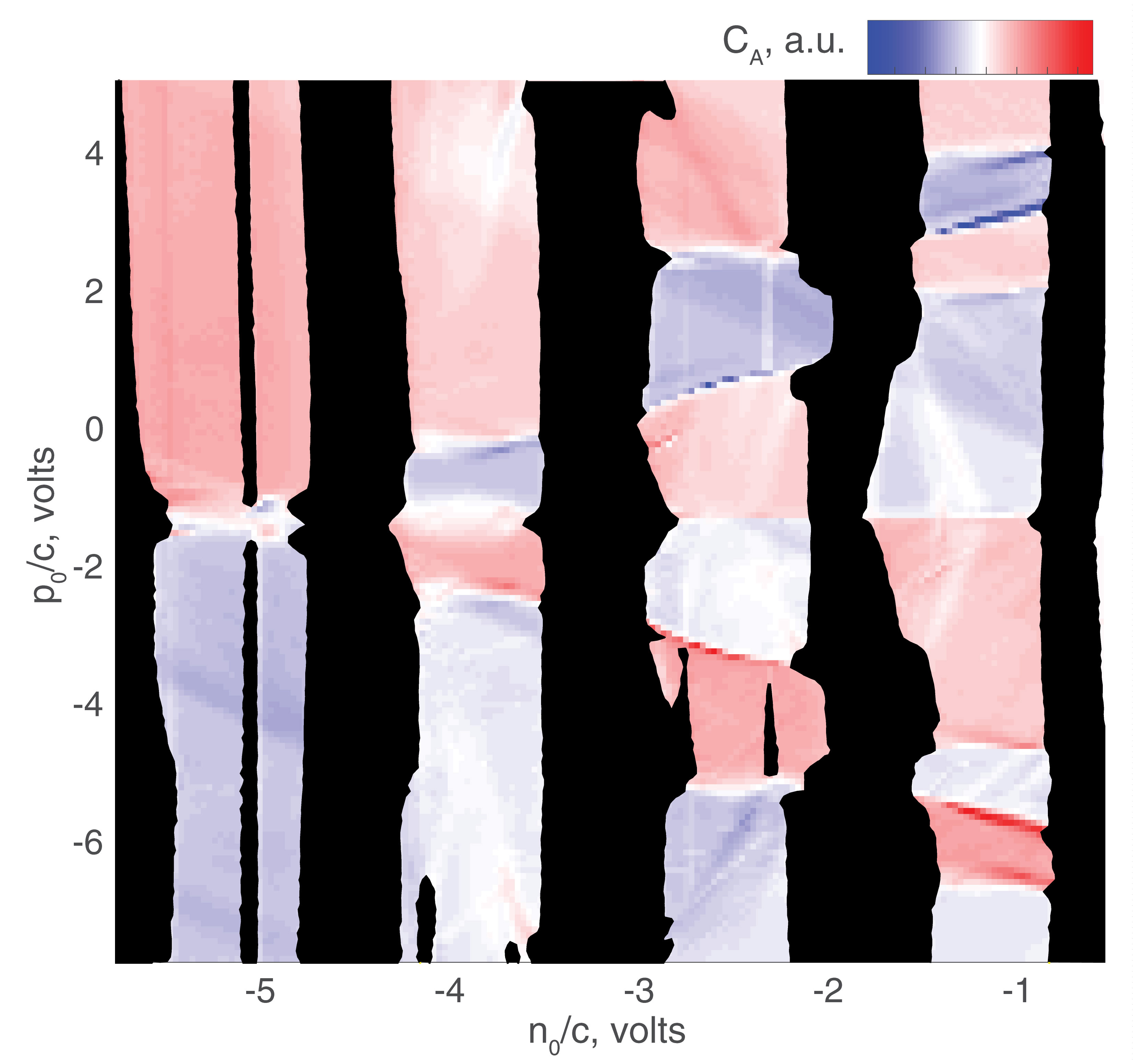}
		\caption{$C_A$ measured in a different device at B=35T for $-4<\nu<0$. The device has nearly identical geometry to that described in the main text, but a larger area, enabling higher resolution measurements. Several fractional quantum Hall features are visible at one third filling of different LLs.}
		\label{seconddevice}
	\end{center}
 \end{figure*}

Dissipation data associated with the data sets in the main text are shown in Figs. \ref{CapAndDis1} and \ref{CapAndDis2}.  $C_A$ data from a different device at high magnetic field is shown in Fig. \ref{seconddevice}.

\clearpage 
\section*{Supplementary references}


\end{document}